\documentclass[a4paper,11pt]{article}
\pdfoutput=1 % if your are submitting a pdflatex (i.e. if you have
\usepackage{jcappub} % for details on the use of the package, please
\usepackage{color}%to be able to use colors
\usepackage{amsmath}%Recomended for mathematical documents (use of: eqref,...)
\usepackage{amsfonts}
\usepackage{graphicx}
\usepackage{amssymb}
\usepackage{float}
\usepackage{comment}
\usepackage{appendix}
\usepackage[dvipsnames]{xcolor}
\definecolor{refs}{RGB}{245,156,74}

\newcommand{\dd}{{\rm d}}

\newcommand{\Lag}{\mathcal{L}}

\newcommand{\mK}{\mathcal{K}}

\newcommand{\Phit}{\tilde{\Phi}}
\newcommand{\Qb}{\bar{Q}}

\newcommand{\Ap}{A^+}
\newcommand{\Am}{A^-}

\newcommand{\be}{\begin{equation}}
\newcommand{\ee}{\end{equation}}
\newcommand{\bea}{\begin{eqnarray}}
\newcommand{\eea}{\end{eqnarray}}

\newcommand{\mpl}{M_{\rm Pl}}

\begin{document}

%\rightline{CERN-TH-2022-204}

\title{Resilience of DBI screened objects and their ladder symmetries}

\author[a]{Jose Beltr\'an Jim\'enez,}
\author[a,b]{Dario Bettoni,}
\author[c]{Philippe Brax.}

\affiliation[a]{Departamento de F\'isica Fundamental and IUFFyM, Universidad de Salamanca, E-37008 Salamanca, Spain.}
\affiliation[b]{Departamento de Matemáticas, Universidad de León,
Escuela de Ingenierías Industrial, Informática y Aeroespacial
Campus de Vegazana, s/n
24071 León.}
\affiliation[c]{Institut de Physique Th\'eorique, Universit\'e Paris-Saclay, CEA, CNRS, F-91191 Gif-sur-Yvette Cedex, France.}

\emailAdd{jose.beltran@usal.es}
\emailAdd{dbet@unileon.es}
\emailAdd{philippe.brax@ipht.fr}

\abstract{Scalar field theories with a shift symmetry come equipped with the  $K$-mouflage (or kinetic screening) mechanism that suppresses the scalar interaction between massive objects below a certain distance, the screening radius. In this work, we study the linear response of the scalar field distribution around a screened (point-like) object subject to a long range external scalar field perturbation for the Dirac-Born-Infeld theory. We find that, for regular boundary conditions at the position of the particle, some multipoles have vanishing response for a lacunar series of the multipole order $\ell$ for any dimension. Some multipoles also exhibit a logarithmic running when the number of spatial dimensions is even. We construct a ladder operator structure, with its associated ladder symmetries, formed by two sets of ladders that are related to the properties of the linear response and the existence of conserved charges. Our results exhibit a remarkable resemblance with the Love numbers properties of black holes in General Relativity, although some intriguing differences subsist.}
\date{\today}

%\keywords{}

\date{\today}
\maketitle
\newpage
\section{Introduction}

The deformability of astrophysical objects has recently received new insights~\cite{Kol:2011vg,Hui:2020xxx,Hui:2021vcv,Pereniguez:2021xcj,BenAchour:2022uqo,Riva:2023rcm,Rai:2024lho}. When considering neutron stars for instance \cite{Hinderer:2007mb}, the Love numbers associated to  their deformation when subject to long range stresses have been shown to be potentially relevant to the determination of the equation of state of the nuclear matter inside these stars. These numbers could eventually be measured using gravitational waves~\cite{Flanagan:2007ix}. In the case of black holes (BH), their structure is far more rigid and it has been shown that for Schwarzschild and Kerr BH in four dimensions~\cite{Binnington:2009bb,LeTiec:2020bos,Charalambous:2021mea}, the Love numbers vanish. This has been recently related to certain algebraic properties of the BH perturbation equations where ladder symmetries and associated conserved currents imply the vanishing of the deformability coefficients~~\cite{Hui:2020xxx,Hui:2021vcv,BenAchour:2022uqo,Riva:2023rcm,Rai:2024lho}. It is then a pertinent question to ask whether other (non-gravitational) interactions show analogous properties for certain systems that could help   clarifying further the underlying reason for their existence.

In a previous study~\cite{Jimenez:2022mih}, we have analysed non-trivial electric configurations sourced by point-like charges for a class of non-linear electrodynamics in four spacetime dimensions. These theories have a built-in  screening mechanism as exemplified by the paradigmatic Born-Infeld theory originally introduced to regularise charged particles' self-energies~\cite{Born:1934gh,Born:1933pep}, thus possibly making it the first instance of a screening mechanism. In this case, the electric field of a point charge saturates to a constant value deep inside a region determined by the screening scale, while outside this region the electric field acquires its usual Maxwellian behaviour. As such, the screened region admits an appealing interpretation as a soft object concentrating in a finite region all the electric energy of the configuration.\footnote{These properties have been exploited in a scenario with charged dark matter whose gauge mediator is described by Born-Infeld's theory so the effects on structure formation only affect the late-time evolution and could potentially account for the Hubble tension \cite{BeltranJimenez:2020tsl,BeltranJimenez:2020csl,BeltranJimenez:2021imo}.} When applying electromagnetic perturbations characterised by their angular momentum $\ell$, we unveiled that the Born-Infeld theory is remarkably resilient among non-linear electrodynamics as  its deformability coefficients, defined as the linear response coefficients to the external excitation, vanish for odd $\ell\ge 3$ for the electric polarisabilities and even $\ell \ge 0$ for the magnetic susceptibilities~\cite{Jimenez:2022mih}. It is also noteworthy that the  Born-Infeld theory admits a ladder operator structure and their associated ladder symmetries that can be associated to the vanishing of the deformability coefficients and the existence of a tower of conserved charges~\cite{Jimenez:2022mih}. These results show certain resemblance with similar findings for BHs and de Sitter spacetime (see e.g. \cite{Lagogiannis:2011st,Compton:2020cjx,Hui:2021vcv,BenAchour:2022uqo,Berens:2022ebl}), although some intriguing differences subsist. For instance, the ladder structure for Born-Infeld  comprises two different ladders, instead of one single ladder as for BHs and de Sitter, that act in a combined manner to reach all the multipoles. Furthermore, the lowest multipole ladder operators have  non-trivial kernels in the space of solutions that have not been observed in the mentioned gravitational systems.

Motivated by the intriguing results found for Born-Infeld electromagnetism, in this work we will turn our attention to the simpler case of scalar field theories featuring a $K$-mouflage screening ~\cite{Babichev:2009ee,Brax:2012jr,Burrage:2014uwa,Brax:2014wla} (the scalar equivalent of the mentioned screening for non-linear electrodynamics) and, in particular, to the Dirac-Born-Infeld (DBI) theory \cite{Alishahiha:2004eh},\footnote{More precisely, we will focus on the shift-symmetric version of DBI that we will introduce below and can be interpreted as a super-fluid. The screening mechanism for DBI was dubbed D-BIonic screening in \cite{Burrage:2014uwa} (see also \cite{Dvali:2010jz}).} which is the scalar analogue of Born-Infeld electromagnetism. $K$-mouflaged theories are described by non-linear kinetic interactions\footnote{We should add that such kinetic interactions only contain first order derivatives. Allowing for second order derivatives would result in Galileon-like theories \cite{Nicolis:2008in} and the equivalent screening mechanism is then called the Vainshtein mechanism.} and they commonly feature a conformal coupling of the scalar to matter that gives rise to screened solutions in the presence of point-like matter sources. Viewed from far away, the field distribution around such screened objects resembles a soft object whose deformability will be the main scope of this work. We will carry out an analysis of the static linear perturbations around screened objects analogous to that in \cite{Jimenez:2022mih}, but we will extend it to arbitrary dimension. As we will show, most of the results found for Born-Infeld electromagnetism are recovered for its scalar DBI relative. In particular, we find multipoles with vanishing linear response and an intriguing two-ladder structure. Going to arbitrary dimension, we will reveal that the precise structure of the ladders depends on the dimension but, quite remarkably, we will still find that  the vanishing of the deformability coefficients for some multipoles occurs for any dimension. A novel feature that we will find by going to arbitrary dimension is that some multipoles have a logarithmic running of the linear response, although this only occurs when the number of spatial dimensions is even.

The paper is organised as follows: In  section \ref{sec:bkg}
we recall facts about $K$-mouflage theories. In section \ref{sec:linear} we  analyse the linear response of the screened field distribution around a point mass. In particular, we focus on the vanishing deformability coefficients in the DBI case. In section \ref{sec:ladder}, we relate the vanishing of the deformability to the existence of ladder operators exchanging the modes of the perturbation theory around the screened background. In section \ref{Sec:existence}, we analyse how the ladder operators appear in general and how particular is the DBI case. We will conclude with a discussion on our results in Sec. \ref{Sec:Discussion}. We will finish with Appendix \ref{Appedix} on the underlying supersymmetric quantum mechanics of the deformability problem. 
%%%%%%%%%%%%%%%%%%%%%%%%%%%%%%%%%%%
\section{K-mouflage theories}
\label{sec:bkg}
Throughout this work we will be mostly interested in the DBI theory. However, most of the expressions for the background and perturbations of the scalar field can be worked out for a generic shift-symmetric theory without much effort. Moreover, we will work in $d$ spatial dimensions for the sake of generality in order to analyse how our results for the linear response depend on the number of dimensions. Let us recall that GR black holes have vanishing Love numbers only in four dimensions ~\cite{Binnington:2009bb,LeTiec:2020bos,Charalambous:2021mea}. An analogous result has also been found for point-like charged particles within Born-Infeld electromagnetism \cite{Jimenez:2022mih} in 3+1 dimensions where the vanishing happens for a lacunar series of multipoles. Given the resemblance of the perturbations of DBI and the axial sector of BI, our results will elucidate if $3+1$ dimensions exhibit some special properties for these systems similarly to what happens for black holes.

%%%%%%%%%%%%%%%%%%%%%%%%%%%%%%%%%%%
\subsection{The model}
Let us then consider the action in $D=d+1$ dimensions for a scalar field
\be 
\label{eq:action_general}
\mathcal{S}= \Lambda^{D}\int \dd^{D}x \;\mK(X)+ \mathcal{S}_{\text{m}}[\psi,g^J]\,,
\ee
with $\Lambda$ the energy scale controlling the non-linearities, $\mathcal K$ a generic function of the (dimensionless) kinetic term 
\be
X=-\frac12 \frac{g^{\mu\nu}\partial_\mu\phi\partial_\nu\phi}{\Lambda^{D}}\,
\ee
 and $\psi$ collectively refers to matter fields coupled via the Jordan metric $g_{\mu\nu}^J$ conformally related to the spacetime metric. The Klein-Gordon equation derived from action \eqref{eq:action_general} for a static and spherically symmetric (pressureless) matter configuration described by a density $\rho(r)$ can be written as
\be 
\frac{1}{r^{d-1}}\frac{\dd}{\dd r} \left[ r^{d-1} \mK_X(X) \frac{\dd\phi}{\dd r}\right]= g \rho(r)\,,
\ee
where $g$ measures the strength of the scalar-matter interaction. We can integrate the above equation over a $d$-dimensional sphere containing the mass distribution to obtain
\begin{equation}
    \mK_X \phi'(r) = \frac{g}{S_{d-1}}\frac{M}{r^{d-1}}\,,
\end{equation}
with $S_{d-1} = \frac{2\pi^{d/2}}{\Gamma(d/2)}$ the surface of the unit radius $d$-dimensional sphere and $M=\int \dd^d x\rho$
the mass of the distribution. We will be mainly interested in point-like particles, so $M$ will simply be the corresponding mass. 

Let us now specialize to the case of the $d$-dimensional DBI Lagrangian
\be
\mathcal K = \sqrt{1+2X}-1\,.
\label{Eq:DBI_Lag}
\ee
In this case we get 
\be
\frac{\partial_r\phi}{\Lambda^{(d+1)/2}} = \frac{1}{\sqrt{1+\left(\frac{r}{r_s}\right)^{2(d-1)}}}\,,
\ee
where we have introduced the screening radius $r_s$
\be
r_s^{d-1} = \frac{Q}{S_{d-1}}\left(\frac{\Lambda}{\mpl}\right)^{(d-3)/2} \frac{1}{\Lambda^{d-1}} \,.
\ee
These theories are such that the energy of the field configuration generated by a point charge remains finite as well as the gradient of the scalar field that goes as $\partial_r\phi\simeq\Lambda^{(d+1)/2}$ as we approach the origin. Other choices for {\cal K} also lead to the existence of screening. In general though the energy generated by a point charge is not necessarily finite and the gradient of the scalar field generically diverges near the origin. For instance, if the $K$-essence theory has a behavior near the origin given by $\mK\sim X^n$ with $n$ some positive constant, the gradient of the scalar field goes as $\partial_r\phi\propto r^{-\frac{2}{2n+1}}$, that diverges, except for $n\to\infty$, in which case the DBI result is recovered.  This improved regular behaviour of the scalar field at the origin for DBI will play an important role as we will see below.

After deriving the relevant analytical form for the static background configuration, we will turn our attention to the analysis of the static perturbations around such a solution that will be the core of this work.

\subsection{Static perturbations}
\label{sec:pert}
%%%%%%%%%%%%%%%%%%%%%%%%%%%%%%%%%%%
%%%%%%%%%%%%%%%%%%%%%%%%%%%%%%%%%%%%

Let us now focus on perturbation over the static spherically symmetric background derived in the previous section. We introduce the static field perturbation $\varphi$ as
\begin{equation}
    \phi = \bar\phi(r) + \varphi(r,\vec{\theta})\,,
\end{equation}
where $\vec{\theta}$ are coordinates in the unit sphere that collectively represent the angular variables. The Klein-Gordon equation for the perturbation is 
\begin{equation}
\label{eq:perturbation}
\frac{1}{r^{d-1}}\partial_r\left[r^{d-1} \mK_X\,c_r^2 \partial_r\varphi\right]+\frac{\mK_X}{r^2}\nabla^2_{\Omega}\varphi = 0\,,
\end{equation}
where $\nabla^2_\Omega$ is the Laplacian on the $(d-2)$-dimensional sphere and we have introduced the (radial) propagation speed
\begin{equation}
    c_r^2 = 1 + \frac{2X\mK_{XX}}{\mK_X}\,.
\end{equation}
Equation \eqref{eq:perturbation} suggests introducing the field
\be
\Phi\equiv r^{d-1}\mK_Xc_r^2\varphi'
\label{eq:defPhi}
\ee
so that it takes the form
\be
\Phi'+\mK_Xr^{d-3}\nabla^2_\Omega\varphi=0.
\label{Eq:eqPhiI}
\ee
If we take the radial derivative of this equation and use again the definition of $\Phi$ we arrive at the equation
\be
\Phi''-\partial_r\log \big(r^{d-3}\mK_X\big)\Phi'+\frac{1}{r^2c_r^2}\nabla^2_\Omega\Phi=0.
\label{Eq:Phi2}
\ee
Let us pause for a moment to make a pertinent comment. The master equation that we have obtained is not for the DBI field perturbation, but it is rather related to its radial derivative. From a physical point of view, this is justified on the grounds of the shift symmetry of the theory. At a more practical level, our choice of variables is also motivated by the variables employed in \cite{Jimenez:2022mih} in the context of BI electromagnetism, whose relation to the DBI scalar theory will be explained in detail in Sec. \ref{Sec:relationtoBI}. In that case, the variable is related to the perturbation of the electric field that is a more physical (gauge-invariant) quantity than the perturbation of the electric potential. For the DBI theory, the variable $\Phi$ can be related to the perturbation of the force mediated by the scalar field. Evidently, the value of the linear response parameters will depend on the specific variable chosen to describe the system (see the discussion on this point in \cite{Jimenez:2022mih}), but this is just a matter of convention and physics will not depend on it.

\begin{figure}[ht]
    \centering
    \includegraphics[scale=.6]{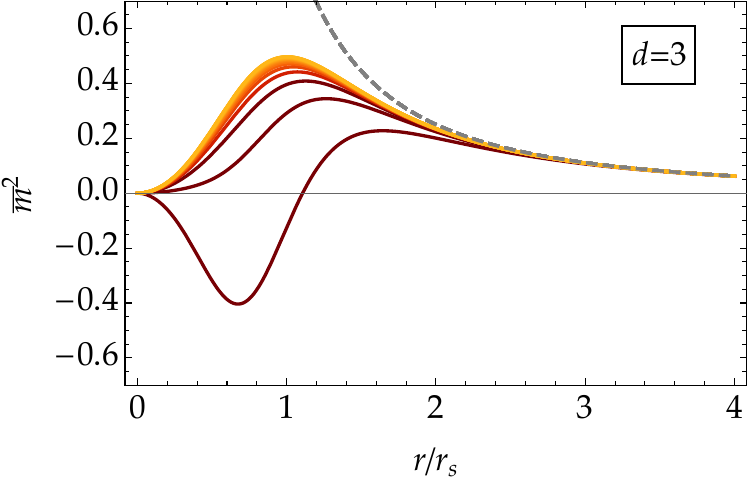}
   \includegraphics[scale=.6]{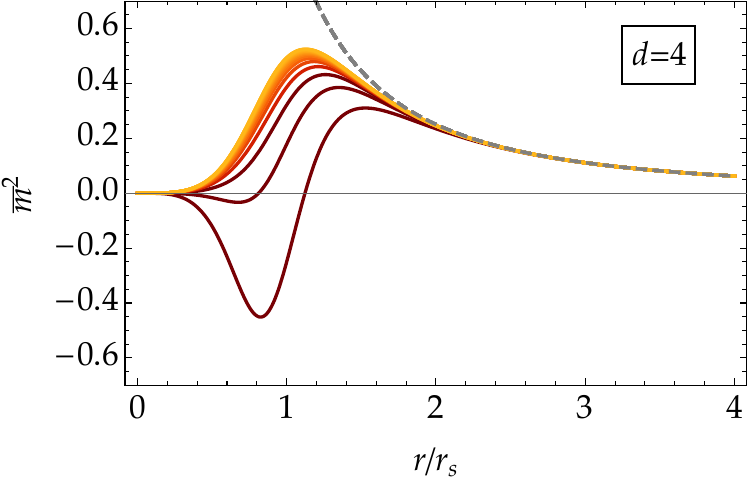}\\
    \includegraphics[scale=.6]{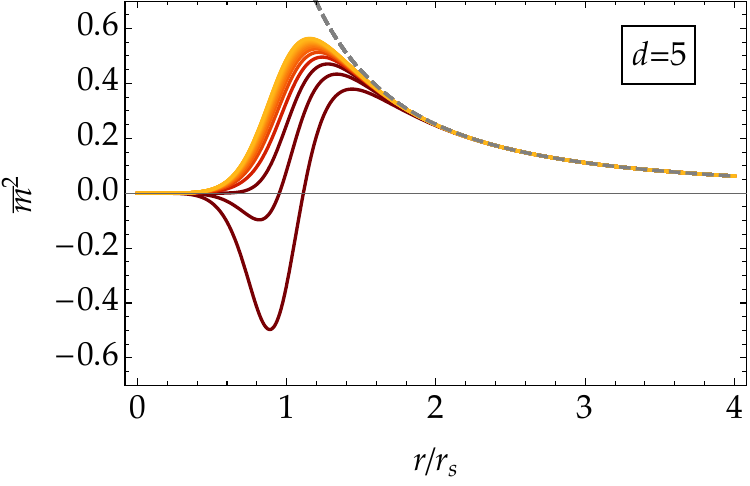}
     \includegraphics[scale=.6]{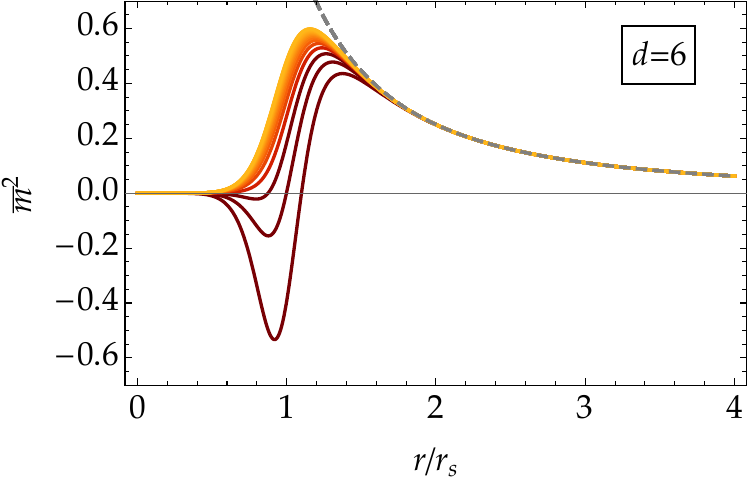}
    \caption{In this plot we show the mass for the perturbations in the DBI theory normalised to $(2\ell+d-3)(2\ell+d-1)/4$ so it goes as $1/r^2$ (gray dashed line) at large distances for all the multipoles. The different colors denote multipoles from 0 (lightest) to 20 (darkest). We can explicitly see the property discussed in the main text that the small $r$ region flattens as we go to higher dimensions. Another interesting feature is the negative values for the lowest multipoles. This could indicate the possibility of forming bound states.}
    \label{fig:MDBI}
\end{figure}

Once we have clarified our choice of variables, let us proceed with the resolution of the the problem. Exploiting the spherical symmetry of the background configuration, we can introduce an expansion in spherical harmonics for the perturbation
\be
\Phi=\sum_{\ell,\vec{m}}\sqrt{r^{d-3}\mK_X}\Phi_{\ell}(r)Y_{\ell,\vec{m}}(\vec{\theta})
\label{eq:multexp}
\ee
where the factor $\sqrt{r^{d-3}\mK_X}$ is introduced to eliminate the friction term in \eqref{Eq:Phi2} and $\Phi_\ell(r)$ are the corresponding multipole components of the perturbations. Due to the spherical symmetry of the background, the perturbations do not depend on $\vec{m}$ so we have dropped it as an index. Plugging the expansion in spherical harmonics into \eqref{Eq:Phi2} we finally obtain the equation
\be
\Phi''_{\ell}-m_\Phi^2\Phi_{\ell}=0,
\label{eq:eqPhim1}
\ee
with the effective mass
\be
m_\Phi^2=\frac{\ell(\ell+d-2)}{r^2c_r^2}+\frac14\left[\partial_r\log\left(r^{d-3}\mK_X\right)\right]^2-\frac12\partial_r^2\log\left(r^{d-3}\mK_X\right)\,.
\label{eq:m2}
\ee
These expressions are valid for any theory characterised by a function ${\cal K}$, although in this work we will focus on the DBI theory, whose masses are shown in Fig. \ref{fig:MDBI} for several multipoles and dimensions. The asymptotic masses at large distances $r\to\infty$ are given by
\be
m_\Phi^2\simeq\frac{(2\ell+d-3)(2\ell+d-1)}{4r^2}\,
\ee
as can be immediately seen from \eqref{eq:m2} and taking into account that $\mK_X\to1$ and $c_r^2\to1$. On the other hand, the masses vanish at the position of the particle. In fact, near the origin we find $m_\Phi^2\propto r^{2(d-2)}$ so the larger the number of dimensions, the faster it goes to zero.

%------------------------------------------------------------
%------------------------------------------------------------

\subsection{Relation to Born-Infeld electromagnetism}

Before proceeding to the analysis of the solutions of the perturbations, we will briefly comment on the close relation between the DBI scalar theory described by \eqref{Eq:DBI_Lag} and the Born-Infeld theory of non-linear electromagnetism that, in 4 dimensions, is described by
\begin{align}
\mathcal{L}_{\text{BI}}=\Lambda^4\left[1-\sqrt{\det\left(\eta_{\mu\nu}+\frac{1}{\Lambda^2}F_{\mu\nu}\right)}\;\right]
=\Lambda^4\left[1-\sqrt{1-\frac{2Y}{\Lambda^4}+\frac{Z^2}{\Lambda^8}}\;\right]\,,
\end{align}
with $Y=-\frac14F_{\mu\nu}F^{\mu\nu}$ and $Z=-\frac14F_{\mu\nu}\tilde{F}^{\mu\nu}$ the two independent Lorentz invariants constructed out of the electromagnetic field strength $F_{\mu\nu}=\partial_{\mu}A_\nu-\partial_\nu A_\mu$. This theory exhibits a kinetic screening mechanism analogous to the scalar $K$-mouflage that was in turn the feature originally invoked by Born and Infeld to regularise the energy of a point-like particle. 

The static perturbations of the electric field generated by a spherical point-like screened object have been studied in \cite{Jimenez:2022mih} where the equivalents of equation \eqref{eq:eqPhim1} for both the polar and the axial sectors have been worked out. For the polar sector, the effective mass exactly coincides with \eqref{eq:m2} for $d=3$. This should not come as a surprise as expanding around a purely electric background and only considering static polar perturbations is equivalent to perturbing the DBI system. To be more explicit, if we consider purely electric perturbations around an electric background, i.e.,  $\vec{E}=\vec{E}_0(r)+\delta\vec{E}(\vec{x})$ and $\vec{B}=0$, then the two independent Lorentz invariants read $Y=\frac12 [\vec{E}_0(r)+\delta\vec{e}(\vec{x})]^2$ and $Z=0$. For this configuration, there is an equivalence between non-linear electromagnetism and K-mouflage theories because the Lagrangian describing this configuration in a non-linear electrodynamics with Lagrangian $\mathcal{L}(Y,Z)$ is simply $\mathcal{L}(Y,0)$. Since the Bianchi identities tell us that we can write the electric field as $\vec{E}=\nabla\phi_e$ with $\phi_e$ the scalar electric potential, we can write $\mathcal{L}(Y,0)=\mathcal{L}(\frac12(\nabla\phi_e)^2,0)$ that exactly reproduces the Lagrangian of a K-mouflage theory in a static configuration upon the identification $\mathcal{L}(Y,0)\to\mathcal{K}(X)$ with $Y\to-X$ and $\phi_e\to\phi$. The equivalence can be shown at the level of the field equations as well since, in the static and electric configuration, the electric potential for a certain charge distribution $\rho_e$ is governed by the equation
\be
\nabla\cdot(\mathcal{L}_Y\nabla\phi_e)=\rho_e
\ee
which, again, exactly coincides with the equation for the shift-symmetric $K$-essence field under the considered conditions and, in particular, for the DBI and Born-Infeld theories. Notice that the equivalence holds at the full non-linear order as long as we have a vanishing magnetic field. For a generic non-linear electrodynamics the equivalence does not hold because a static electric field could source a static magnetic field, thus making inconsistent the considered purely electric configuration. This can be seen by writing the full set of equations for a static configuration, given by
\bea
\nabla\cdot\Big(\Lag_Y\vec{E}+\Lag_Z\vec{B}\Big)=\rho\,,\\
\nabla\times\Big(\Lag_Y\vec{B}-\Lag_Z\vec{E}\Big)=0\,,
\label{eq:eomEB}
\eea
where we see that $\vec{B}$ is sourced by the electric field provided $\Lag_Z\neq0$. However, if we impose parity then $\Lag_Z$ vanishes in a pure electric configuration so it is consistent to keep $\vec{B}=0$ and the equivalence to the DBI theory holds. The discussed equivalence permits to unveil a symmetry that arises in BI electromagnetism for static electric configurations as the analogous of the DBI symmetry. More explicitly, the Born-Infeld Lagrangian in a static electric configuration is
\be
\Lag_{\text{BI}}=\Lambda^4\left[1-\sqrt{1-\frac{\vert\nabla\phi_e\vert^2}{\Lambda^4}}\;\right],
\ee
that, under the transformation
\be
\delta\phi_e=c+\vec{v}\cdot\vec{x}-\frac{1}{\Lambda^2}\phi_e\vec{v}\cdot\nabla\phi_e\,,
\ee
changes as the following total derivative:
\be
\delta \Lag_{\text{BI}}=-\nabla\cdot\left[{\Lambda^2\phi_e\sqrt{1-\frac{\vert\nabla\phi_e\vert^2}{\Lambda^4}}\vec{v}}\right].
\ee
While in the scalar DBI theory this symmetry is reminiscent of a 5 dimensional Lorentz invariance, the Born-Infeld electromagnetism in the static electric configuration only possess it as an accidental symmetry. However, it is interesting to notice that Born-Infeld electromagnetism could be selected by imposing the existence of the symmetry in those backgrounds. This is of course the equivalent of the symmetry selecting DBI as argued in \cite{Pajer:2018egx}. It would be interesting to explore this connection further, but we will not proceed along these lines here.

After this brief digression on the relation with Born-Infeld electromagnetism, we will now turn our attention to the analysis of the perturbation equations \eqref{eq:eqPhim1} for the different multipoles and, in particular, to the linear response to external stimuli.

\section{Linear responses}
\label{sec:linear}

 In this section we will tackle the main goal of this work, namely, solving the perturbation equations for the static sector that will enable us to compute the linear response. For the solution of the perturbations we need to implement appropriate boundary conditions that we will set at the position of the particle and in the asymptotic region. For our purposes, the asymptotic boundary condition is not important because it will simply correspond to the amplitude of the external perturbation. At the origin, we will exploit the improved behaviour of the DBI theory that permits to impose a regular boundary condition at the position of the particle as we will see. This boundary condition presents some rigid properties that will translate into the vanishing of some coefficients of the linear response for some multipoles. We will also see that some dimensions exhibit a logarithmic running of the linear response. Let us then commence by obtaining the solutions that are regular at the origin.

\subsection{Regular solution at the origin}
\label{Sec:relationtoBI}

In order to solve the equation of the perturbations, we will first transform the equation into a more familiar form. Inspired by \cite{Jimenez:2022mih} (that took inspiration from \cite{Compton:2020cjx,Hui:2020xxx}), we introduce a new radial variable $z\equiv-x^{2(d-1)}$ and re-scale the multipole perturbation as
\be
\tilde{\Phi}_\ell(z)\equiv \left(1-z\right)^{1/4}\Phi_\ell\,,
\ee
which turns Eq. \eqref{eq:eqPhim1} into a hypergeometric equation
\be
z(1-z)\tilde{\Phi}_\ell''(z)+\frac{2d-3-(d-2)z}{2(d-1)}\tilde{\Phi}_\ell'(z)+\frac{(\ell+d-1)(\ell-1)}{4(d-1)^2}\tilde{\Phi}_\ell(z)=0,
\label{eq:HyperPhi}
\ee
with parameters 
\be
a=-\frac12\left(1+\frac{\ell}{d-1}\right),\quad b=\frac{\ell-1}{2(d-1)},\quad c=1-\frac{1}{2(d-1)},
\label{eq:hypparameters}
\ee
that satisfy $c-a-b=\frac32$. Thus, the space of solutions can be spanned by the two independent solutions$\;_2F_1(a,b,c;z)$ and $z^{1-c}\;_2F_1(1+a-c,1+b-c,2-c;z)$. Since $c$ is never an integer (except for $d=\infty$, but we do not consider this strict case), these two solutions are non-degenerate for arbitrary dimension and the general solution can be written as
\be
\tilde{\Phi}_\ell=A_\ell\;_2F_1(a,b,c;z)+B_\ell (-z)^{1-c}\;_2F_1(1+a-c,1+b-c,2-c;z)\,,
\ee
with $A_\ell$ and $B_\ell$ integration constants and we have taken into account that $z$ is negative. The solution for our original variable $\Phi_\ell$ can be written as
\be
\Phi_\ell=\frac{1}{(1-z)^{1/4}}\Big[A_\ell\;_2F_1(a,b,c;z)+B_\ell\, x\;_2F_1(1+a-c,1+b-c,2-c;z)\Big].
\ee

Equipped with the general solution for the perturbations, we can proceed to obtain the linear response to external scalar stimuli. For that, we need to impose appropriate boundary conditions. A natural place to impose a boundary condition is the origin. By using the expansion of the hypergeometric function near the origin $_2F_1(a,b,c;z)\simeq 1+\frac{ab}{c}z$, we find that the general solution for small $z$ reads\footnote{In the subsequent expressions we will use both $x$ and $z$ for the expressions to look more amicable, but it should be understood that there is only one radial variable.}
\be
\Phi_\ell\simeq A_{\ell}\left[1-\left(\frac{(\ell+d-1)(\ell-1)}{2(2d-3)(d-1)}-\frac14\right)z\right]+B_\ell x \left[1-\left(\frac{\ell(\ell+d-2)}{2(2d-1)(d-1)}-\frac14\right)z\right].
\ee
Given the regular nature of the DBI theory at the origin for the background configuration, we will require regularity at the origin also for the perturbations and the tightest constraint  comes from the angular gradients of the scalar field. Indeed, the gradient of the DBI field perturbation can be expressed as
\be
\nabla\varphi=\varphi'\hat{r}+\frac{1}{r}\nabla_\Omega\varphi=\sum_{\ell,\vec{m}}\left[\varphi_\ell'(r)Y_{\ell,\vec{m}}\hat{r}+\frac{\varphi_\ell(r)}{r}\nabla_\Omega Y_{\ell,\vec{m}}\right],
\ee
where we have decomposed $\varphi=\sum_{\ell,\vec{m}}\varphi_\ell(r) Y_{\ell,\vec{m}}(\vec{\theta})$. This expression shows that we need $\varphi_\ell$ to vanish at least as fast as $\sim r$ as we approach the origin for the angular components to remain regular. From the definition of $\Phi$ given in \eqref{eq:defPhi} and its multipole expansion given in \eqref{eq:multexp} we find
\be
\varphi'_\ell=\frac{\Phi_\ell}{\sqrt{x^{d+1}\mK_X}c^2}\;.
\ee
If we now use that $c_r^2\simeq x^{2(1-d)}=-\frac{1}{z}$ and $x^{d-1}\mK_X\simeq1$ as we approach the origin $x\to0$, we obtain
\be
\varphi'_\ell\simeq\big(A_\ell+B_\ell x\big)x^{2d-3},
\ee
so it remains regular at the origin for any dimension $d>1$, which comprises all the cases of interest to us (let us recall that we are assuming $d\geq3$). In order to obtain the angular gradients, we will use \eqref{Eq:eqPhiI} that gives
\be
\Phi'=-r^{d-3}\mK_X\nabla^2_\Omega\varphi,
\ee
so, taking into account the normalisation of the multipoles and in terms of the variable $x$, we have
\be
\varphi_\ell=\frac{\Big(\sqrt{x^{d-3}\mK_X}\Phi_\ell\Big)'}{\ell (\ell+d-2) x^{d-3}\mK_X}.
\ee
For small $x$, we then find, at leading order in $x$, 
\be
\varphi_\ell\simeq -\frac{A_\ell}{\ell(\ell+d-2)}+\left[\frac{1}{2d-1}-\frac{d-1}{2\ell (\ell+d-2)}\right]B_\ell x^{2d-1}\,.
\ee
This shows that we must impose $A_\ell=0$ to avoid the singular behaviour of the angular gradients $r^{-1}\nabla_\Omega\varphi$ at the origin. Notice that the monopole $\ell=0$, that would appear as singular, can be absorbed into a shift of the background scalar charge. The regular solution then reads
\be
\tilde{\Phi}^{\text{reg}}_\ell=B_\ell (-z)^{\frac{1}{2(d-1)}}\;_2F_1\left(\frac{\ell}{2(d-1)},-\frac{\ell+d-2}{2(d-1)},1+\frac{1}{2(d-1)};z\right)\,,
\ee
or, for our original multipole perturbation,
\be
\Phi^{\text{reg}}_\ell=B_\ell \frac{x}{(1-z)^{1/4}}\;_2F_1\left(\frac{\ell}{2(d-1)},-\frac{\ell+d-2}{2(d-1)},1+\frac{1}{2(d-1)};z\right)\,.
\label{eq:RegSol}
\ee
This concludes the resolution for the multipole perturbations that remain regular at the origin and we can now proceed to compute the linear response to external perturbations, whose amplitude will be encoded in the (irrelevant for us) value of $B_\ell$.

\subsection{Linear response}

In order to obtain the linear response, we expand the obtained regular solution in the asymptotic region $x\to\infty$ where it exhibits a growing mode $\sim x^{\ell+(d-1)/2}$ that represents the external perturbation (i.e. the second boundary condition) and a  decaying tail $\sim x^{-\ell-(d-3)/2}$ describing the response of the particle. The linear response is then defined as the ratio of the coefficients of the decaying and growing modes. We can now use the connection formula
\begin{align}
_2F_1(a,b,c;z)=\frac{\Gamma(b-a)\Gamma(c)}{\Gamma(b)\Gamma(c-a)}(-z)^{-a}\;_2F_1(a,1-c+a,1-b+a;z^{-1})\nonumber\\+\frac{\Gamma(a-b)\Gamma(c)}{\Gamma(a)\Gamma(c-b)}(-z)^{-b}\;_2F_1(b,1-c+b,1+b-a;z^{-1}),
\label{eq:AsympRegSol}
\end{align}
valid whenever $a$, $b$ and $a-b$ are not integer. We will assume for now that they are not integer and we will come back to the integral cases in a subsequent section below. Since $a=\frac{\ell}{2(d-1)}$ is positive and $b=-\frac{\ell+d-2}{2(d-1)}$ is negative for the regular solution \eqref{eq:RegSol}, the first term gives the decaying tail and the second term corresponds to the growing mode so we find the following expression for the linear response:
\be
\alpha_{\ell,d}=\frac{\Gamma\left(\frac{\ell}{2(d-1)}\right)\Gamma\left(-\frac{2\ell+d-2}{2(d-1)}\right)\Gamma\left(\frac32+\frac{\ell}{2(d-1)}\right)}{\Gamma\left(\frac{2d-1-\ell}{2(d-1)}\right)\Gamma\left(\frac{2-d-\ell}{2(d-1)}\right)\Gamma\left(\frac{2\ell+d-2}{2(d-1)}\right)}r_{\text{s}}^{2\ell+d-2}\,.
\label{eq:alpha}
\ee
As one might expect, the scale of the linear response is determined by the appropriate power of the screening radius $r_{\text{s}}$ so we can define the dimensionless linear response
\be
\bar{\alpha}_{\ell,d}\equiv\frac{\alpha_{\ell,d}}{r_{\text{s}}^{2\ell+d-2}}\,.
\ee
In the limit of a large number of dimensions $d\gg  1$, the linear response behaves as
\be
\bar{\alpha}_{\ell,d\gg1}\simeq\frac{d}{\ell}
\ee
which means that, for a given multipole, the linear response grows linearly with the number of dimensions. On the other hand, the limit of high multipoles $\ell\gg1$ yields
\be
\bar{\alpha}_{\ell\gg1,d}\simeq 2^{-\frac{2\ell}{d-1}}.
\ee

\begin{figure}[ht!]
    \centering
    \includegraphics[scale=.7]{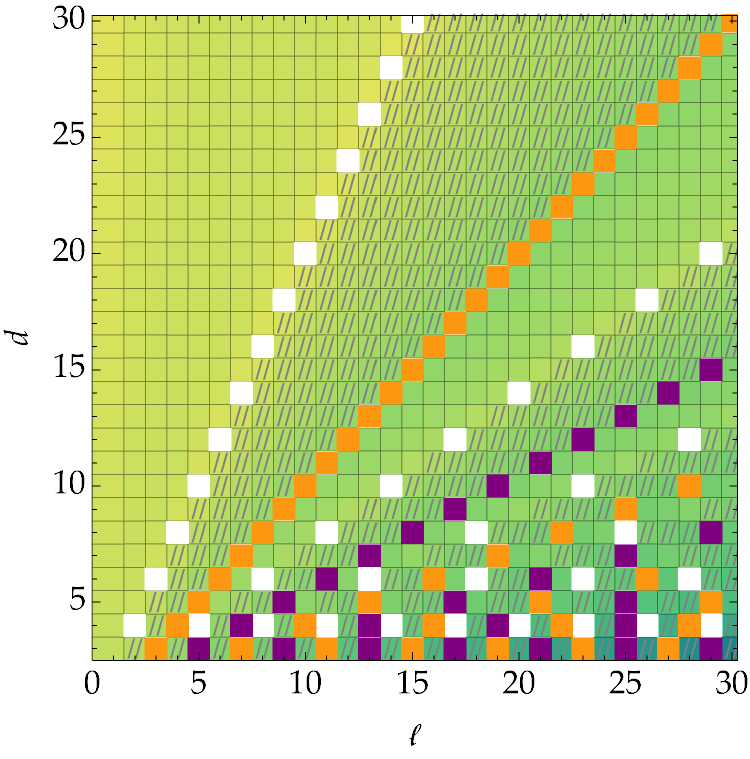}
    \caption{In this figure we show the (absolute) linear response for different values of $\ell$ and $d$ with lighter green meaning larger values. The sign has been denoted by the dashed region that corresponds to negative values of $\alpha_{\ell,d}$. We have also marked in white the cases with the logarithmic running discussed in the text, while the orange and purple squares denote the two series with vanishing response.}
    \label{Fig:alphald}
\end{figure}

In order to obtain the linear response given by \eqref{eq:alpha} we have implicitly assumed that the multipole solutions and the asymptotic expansions are all regular. However, it is well-known that the solutions of the hypergeometric equation can degenerate for some values of the parameters. The expression \eqref{eq:alpha} already warns us that some specific multipoles in some particular dimensions are special because the $\Gamma$ function is singular for non-positive integer values of its argument. We can distinguish two situations, namely, when the denominator becomes singular and when the numerator is singular. We explore these cases next and refer to Fig. \ref{Fig:alphald} for a general overview of the discussed features.

\subsection{Vanishing response}

Let us first turn our attention to the values of $\ell$ and $d$ that correspond to the poles in the $\Gamma$ functions of the denominator of $\alpha$. At face value, this would seem to imply that the linear response vanishes. The poles of these functions occur when the corresponding argument is a non-positive integer. Since the last factor in the denominator has a positive argument for any multipole in $d>2$, we only need to consider the first two terms (the case $d=2$ is special and needs to be treated separately). We thus find two series of multipoles with seemingly vanishing linear response given by
\be
\ell=2(d-1)n+2d-1,\quad\ell=2(d-1)n+2-d,
\ee
with $n$ a non-negative integer, which needs to be larger than $\frac{d-1}{2(d-1)}$ for the second series. In order to understand this result, we can look at the regular solution for these two series of multipoles
\begin{align}
&\ell=2(d-1)n+2d-1\;\Rightarrow\; \Phi^{\text{reg}}_\ell= \frac{ B_\ell\, x}{(1-z)^{1/4}}
\;_2F_1\left(-\frac{3}{2}-n,n+1+\frac{1}{2(d-1)},1+\frac{1}{2(d-1)};z\right),\\
&\ell=2(d-1)n+2-d\;\;\;\Rightarrow\; 
\Phi^{\text{reg}}_\ell= \frac{ B_\ell\, x}{(1-z)^{1/4}}
\;_2F_1\left(-n,n+\frac{2-d}{2(d-1)},1+\frac{1}{2(d-1)};z\right).
\end{align}
The hypergeometric function reduces to a finite polynomial for non-positive integer values of the parameters $a$, $b$. The second series precisely corresponds to having a non-positive integer parameter $a$. This means that the regular solution does not have a decaying mode at $x\to \infty$ and, hence, it explains why we obtain a vanishing linear response. In the case of the first series we instead have that $c-b=-n$ is a non-positive integer, in which case the hypergeometric function also reduces to a polynomial and we again lack a decaying mode that explains the vanishing response for those multipoles.

In both cases, the series comprise multipoles separated by $2(d-1)$ and the only difference is the starting multipoles. The first series starts with the multipole $\ell=2d-1$, while the second series starts at $\ell=d$. We thus see that we obtain resilient multipoles in all dimensions, but they are more scarce as we go to higher dimensions. In fact, the maximum resilience is obtained for $d=3$ where all odd multipoles above the dipole exhibit vanishing deformabilities.

As we will see in the next sections, the two series of multipoles with $\alpha_\ell=0$ can be related to the monopole and the dipole sector via a ladder structure that we will unveil next.

\subsection{Diverging response: Logarithmic running}

The second singular behaviour of the linear response occurs for the multipoles corresponding to poles of the factor $\Gamma\left(-\frac{2\ell+d-2}{2(d-1)}\right)$ in \eqref{eq:alpha} that take place when 
\be
\frac{2\ell+d-2}{2(d-1)}=-k \quad\Rightarrow\quad\ell=(2k-1)\frac{d}{2}+1-k,
\ee
for\footnote{We leave out $k=0$ because it would give a negative $\ell$ that is not physical.} $k=1,2,3,\cdots$. Let us notice that these multipoles only exist in even dimension $d$. In particular, they are absent in $d=3$. The origin of these singular multipoles can be traced to the fact that the analytic continuation of the regular solution at the origin is no longer given by \eqref{eq:AsympRegSol} since the hypergeometric function precisely satisfy $a-b=k$ for the multipoles with diverging $\alpha_{\ell,d}$. Thus, instead of \eqref{eq:AsympRegSol}, we need to use the relation valid for $b=a+m$ with $m=0,1,2,\cdots$ given by\footnote{Let us notice that the diverging multipoles satisfy $a=b+k$ with $k$ a non-negative integer, but we can exploit the symmetry of the hypergeometric function under the exchange of $a$ and $b$ to put it in the form $b=a+m$ and so we can identify $k=m$ in the expressions as long as we identify $a=-\frac{\ell+d-1}{2(d-1)}$ and $b=\frac{\ell}{2(d-1)}$.} 
\begin{align}
\frac{\Gamma(a+m)}{\Gamma(c)}\,_2F_1(a,a+m,c;z)=&\frac{(-z)^{-a-m}}{\Gamma(c-a)}
\sum_{n=0}^\infty \frac{(a)_{n+m}(1-c+a)_{n+m}}{n! (n+m)!}z^{-n}\Big(\log(-z)+h_n\Big)\nonumber\\
&+(-z)^{-a}\sum_{n=0}^{m-1}\frac{(a)_n\Gamma(m-n)}{n!\Gamma(c-a-n)}z^{-n}
\label{eq:AsympRegSol2}
\end{align}
where $h_n$ is given in terms of digamma function $\psi$ as
\be
h_n=\psi(m+n+1)+\psi(n+1)-\psi(a+m+n)-\psi(c-a-m-n).
\ee
The above relation holds for $a\neq0,-1,-2$ and the summation in the second line is to be replaced by zero if $m=0$. Since $m=0$ is excluded from the diverging multipoles and $a$ is never an integer for those multipoles, we do not need to worry about this. In any case, we can see the appearance of a logarithmic correction that modifies the would-be decaying tail. Thus, the diverging response for these multipoles was hinting at the presence of this logarithmic correction. This correction can be interpreted as a classical running of the linear response, i.e., the linear response is no longer a universal quantity, but it depends on the scale at which it is measured. As usual however, once the value is determined at a given scale, its value at any other scale is determined precisely by the said running.

\section{Ladder operators and conserved charges}
\label{sec:ladder}
\subsection{Ladder operators}

The vanishing of certain deformability coefficients can be related to the existence of some ladder symmetries. We will now search for ladder operators following the same procedure as in \cite{Jimenez:2022mih}. We will for the moment leave out the details of the computation and we will defer them to Sec. \ref{Sec:existence} where we will carry out a more systematic and thorough analysis of the existence of ladder operators. A different approach, leading to equivalent results and using supersymmetric quantum mechanics is detailed in Appendix \ref{Appedix}. 

The search for ladder operators consists in introducing the family of Hamiltonians
\be
H_{d,\ell}\equiv z(z-1)\left[z(1-z)\frac{\dd^2}{\dd z^2}+\frac{2d-3-(d-2)z}{2(d-1)}\frac{\dd}{\dd z}+\frac{(\ell+d-1)(\ell-1)}{4(d-1)^2}\right]\,,
\ee
whose kernels coincide with the space of solutions of the multipole equations \eqref{eq:HyperPhi} and, then, finding ladder operators that factorise them. By doing so, we obtain that two ladders are possible for $d>2$, one that connects $\ell$ with $\ell+2(d-1)$ and another one that connects $\ell$ and $d-\ell$. We will refer to them as big and small ladders. Let us start with the big ladder.

\subsubsection{Big ladder}

The big ladder operators are
\bea
A^-_{d,\ell}&=&z(z-1)\frac{\dd}{\dd z}+\frac{\ell+3(d-1)}{2(d-1)}\left[\frac{\ell}{2\ell+3d-4}-z\right],\\
A^+_{d,\ell}&=&-z(z-1)\frac{\dd}{\dd z}+\frac{\ell-1}{2(d-1)}\left[\frac{\ell+3d-4}{2\ell+3d-4}-z\right],
\eea
that factorise the Hamiltonian as
\begin{align}
A^-_{d,\ell}A^+_{d,\ell}=H_{d,\ell}+\varepsilon_{d,\ell}\,,\qquad A^+_{d,\ell}A^-_{d,\ell}=H_{d,\ell+2(d-1)}+\varepsilon_{d,\ell}\,,
\end{align}
with 
\be
\varepsilon_{d,\ell}=\frac{\ell(\ell-1)(\ell+3d-4)(\ell+3d-3)}{4(d-1)^2(2\ell+3d-4)^2}\,.
\ee
This factorisation further yields the following intertwining relations
\be
A^+_{d,\ell}H_{d,\ell}=H_{d,\ell+2(d-1)}A^+_{d,\ell}\qquad\text{and}\qquad A^-_{d,\ell}H_{d,\ell+2(d-1)}=H_{d,\ell}A^-_{d,\ell}.
\ee
The interest of the ladder operators is their ability to connect solutions for different multipoles. If we have $\Phit_\ell$ that satisfies $H_{d,\ell}\Phit_{\ell}=0$, then we can use the raising operator $A^+_{d,\ell}$ to construct $\Phit_{\ell+2(d-1)}\equiv A^+_{d,\ell}\Phi_\ell$ that satisfies 
\be
H_{d,\ell+2(d-1)}\Phit_{\ell+2(d-1)}=H_{d,\ell+2(d-1)}A^+_{d,\ell}\Phi_\ell=A^+_{d,\ell}H_{d,\ell}\Phi_\ell=0,
\ee
i.e., it is a solution of the equation for the multipole $\ell+2(d-1)$. Analogously, the lowering operator $A_{d,\ell}$ connects solutions for the multipoles  $\ell+2(d-1)$ with solutions for $\ell$. This structure would, in principle, allow to connect the lowest $2d-1$ multipoles with all the higher multipoles, i.e., we can construct the solutions for all the multipoles from the first $2d-1$ via the ladder operators. As we will see shortly, there is an obstruction for this construction to work in the suggested clear manner. 

Before discussing this obstruction, let us notice an important property of these ladder operators and how they connect solutions of different multipoles. Near the origin, both ladder operators take the approximate form $A^{\pm}_{d,\ell}\vert_{z\to0}\simeq \pm z\partial_z+c_\pm$ which shows that the ladder operators do not change the behaviour of the solutions at the origin. This means that they preserve the regularity character of the solutions, i.e., they connect regular solutions to regular solutions and singular solutions to singular solutions. This is a remarkable property that will be very useful because if we start from a regular solution, we are guaranteed that all the solutions for the higher multipoles obtained via the ladder operators will also be regular. In the asymptotic region $z\to-\infty$, the ladder operators are instead $A^{\pm}_{d,\ell}\vert_{z\to-\infty}\simeq \pm z^2\partial_z+C_\pm z$ so they essentially act by raising one power of $z$ the asymptotic form of the multipoles.  This means that the ladder operators will preserve the property of lacking a decaying tail so that the multipoles with vanishing linear response will be connected via the ladder. On the other hand, the multipoles with a logarithmic running will also be connected via the ladder because the logarithmic growth will also be preserved.

Let us now notice that both the monopole and the dipole have $\varepsilon_{d,0}=\varepsilon_{d,1}=0$. This represents an obstruction for the complete connection of the monopole and dipole spaces of solutions with the corresponding spaces of solutions of multipoles $2(d-1)$ and $1+2(d-1)$ that would-be connected via the ladder operators. For instance, if we consider a monopole solutions $\Phit_0$ and we construct its relative $\Phit_{2(d-1)}\equiv A^+_{d,0}\Phit_0$, then we cannot connect it back to a monopole solution because this would be done with $A^-_{d,0}$, but we have $A^-_{d,\ell}A^+_{d,\ell}\Phit_{d,0}=H_{d,0}\Phit_{d,0}=0$. Thus, instead of taking it back to the regular solution for the monopole, it is annihilated by the lowering operator. Likewise, the space of monopole solutions constructed from the dipole space of solutions $\Phit_1$ via the lowering operator as $\Phit_0\equiv A^-_{d,0}\Phit_{d,1}$ is in the kernel of $A^+_{d,0}$ because $A^+_{d,0}A^-_{d,0}\Phit_{d,1}=H_{d,1}\Phit_{d,1}=0$.

We will finish our discussion on the big ladder by noticing that the jump coincides with the jump of the multipoles the exhibit vanishing linear response. This means that the multipoles with $\alpha_\ell$ are connected via the big ladder.

\subsubsection{Small ladder}
The small ladder is generated by the operators
\bea
a^-_{d,\ell}&=&z(z-1)\frac{\dd}{\dd z}+\frac{\ell-2d+1}{2(d-1)}\left[\frac{\ell+d-2}{d-2\ell}+z\right],\\
a^+_{d,\ell}&=&-z(z-1)\frac{\dd}{\dd z}+\frac{\ell+d-1}{2(d-1)}\left[\frac{\ell-2(d-1)}{d-2\ell}+z\right].
\eea
Let us notice that this small ladder is ill-defined in even dimensions for the multipole $\ell=d/2$. Since the small ladder would connect such a multipole with itself, this feature might have been anticipated.

These ladders show a gap for $d>3$ and only in $d=3$ we can reach all the multipoles starting from the monopole and the dipole. This gap should be related to the scarcer distribution of multipoles with vanishing deformability.

For the small ladder we have
\begin{align}
a^-_{d,\ell}a^+_{d,\ell}=H_{d,\ell}+\delta_{d,\ell}\,,\qquad a^+_{d,\ell}a^-_{d,\ell}=H_{d,d-\ell}+\delta_{d,\ell}\,,
\end{align}
with 
\be
\delta_{d,\ell}=\frac{(\ell+d-1)(\ell+d-2)(\ell-2d+1)(\ell-2d+2)}{4(d-1)^2(2\ell-d)^2}\,.
\ee
These quantities will vanish for $\ell=2d-1$ and $\ell=2d-2$ so for these multipoles we might expect having an obstruction for the connection of different multipole solutions via the small ladder similar to the big ladder situation. However, for $d\geq3$, the small ladder never connects modes with $\delta_\ell=0$ because $d-\ell$ is always smaller than $2d-2$ (and, hence, smaller than $2d-1$) so we do not need to worry about the incompleteness of the mapping between the multipoles connected by the small ladder.

\begin{figure}
\includegraphics[width=\textwidth]{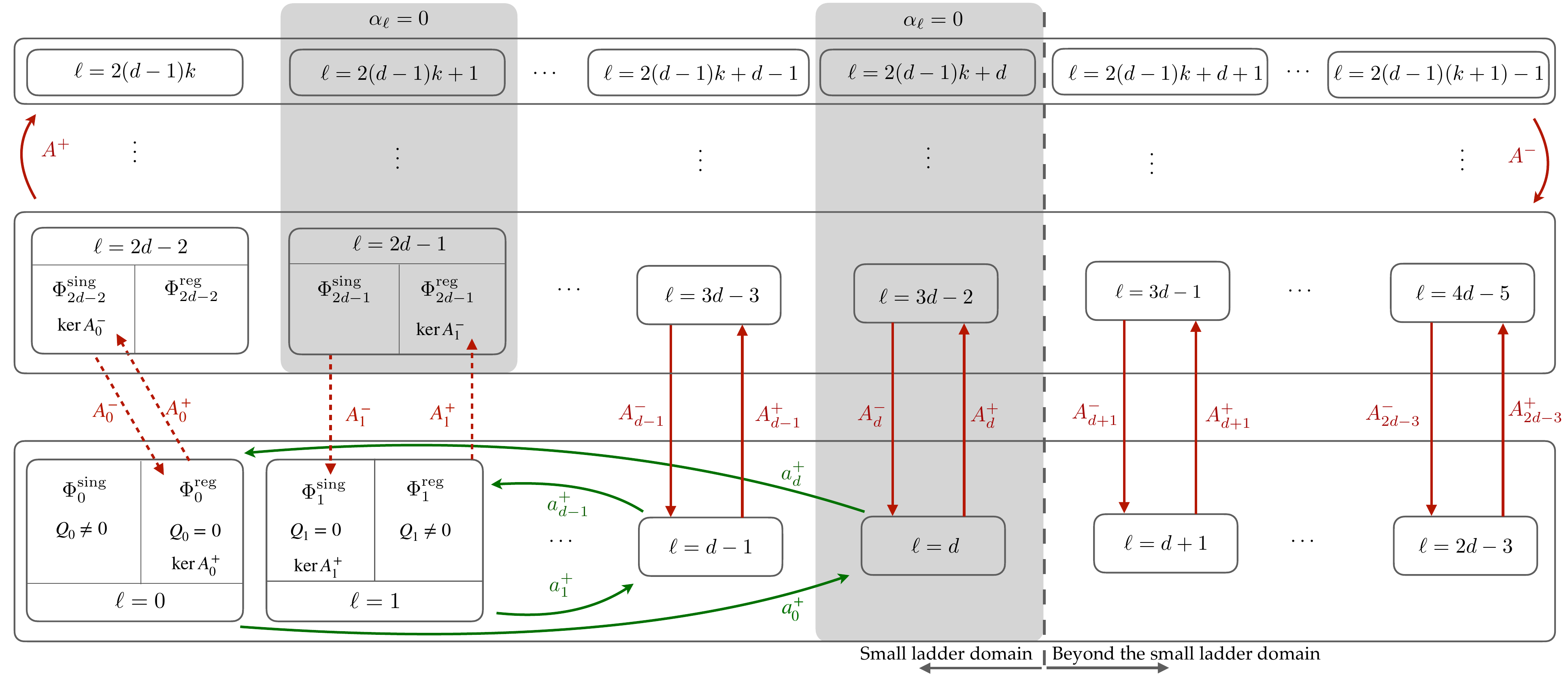} 
\vspace{-0.7cm}
\caption{In this Figure we show the ladder structure for the perturbations of the DBI screened system. The multipoles with vanishing linear response are indicated by the gray-shadow.
} 
\label{Fig:Ladders}
\end{figure}

We can now establish the full ladder structure and the corresponding links between the different multipoles via both ladders (see Fig. \ref{Fig:Ladders}). Firstly, the small ladder permits to connect the first multipoles up to  $\ell=d$ multipoles that could then be grouped together. On the other hand, the big ladder allows to climb up in steps of $2(d-1)$ multipoles so we can arrange them into floors of $2(d-1)$ multipoles with the ground floor containing the multipoles up to $\ell=2d-3$  that could then be connected via the big ladder. Let us notice that, with this arrangement, the small ladder corresponds to an automorphism between the first $d$ multipoles of the ground floor. A remarkable coincidence occurs in $d=3$ for which the last multipole of the ground floor $\ell=2d-3$ coincides with the multipole $d$ that is connected to the monopole via the small ladder. This means that, for $d=3$, the ground floor precisely corresponds to the multipoles that are grouped together by means of the small ladder. For higher dimensions, this is no longer true and the ground floor contains multipoles that lie beyond the domain of the small ladder. 

\subsection{Conserved currents}
After constructing the ladder structure exhibited by the perturbations, we will proceed to the construction of a hierarchy of conserved charges that can be constructed by employing the unveiled ladders. The starting point is to notice that the monopole and the dipole have conserved charges generated by 
\begin{align}
\mathcal{Q}_0[\Phi]=\frac{z^{\frac{2d-1}{2(d-1)}}}{\sqrt{1-z}}\frac{\dd}{\dd z}\left[\frac{\Phi}{z^{\frac{1}{2(d-1)}}}\right]\,,\qquad
\mathcal{Q}_1[\Phi]=\frac{z^{\frac{2d-1}{2(d-1)}}}{\sqrt{1-z}}\frac{\dd \Phi}{\dd z}\,,
\end{align}
as can be seen by simple inspection. In fact, both equations can be written as $\frac{\dd\mathcal{Q}_{0,1}[\Phit]}{\dd z}=0$, thus explicitly showing the existence of the conserved charges \footnote{The conserved charges for higher multipoles are defined by first acting with the ladder operators to go back to the monopole or the dipole. The resulting state has a charge ${\cal Q}_{0,1}$ depending on whether the ladder lands at the monopole or the dipole. This is the charge of the considered multipole state.  The application of the inverse ladder allows one to bring back the state to the appropriate multipole level. Schematically we have 
${\cal Q}_\ell [\Phi_\ell]= {\cal Q}_{0,1}[A^- \Phi_\ell]$ where $A^-$ is the appropriate ladder operator. Notice that the existence of these conserved currents is not guaranteed at all levels because of the "gap" in the small ladder structure between $\ell=d$ and $\ell= 2d-3$, see figure 3. This gap closes only when $d=3$. }.It turns out to be very intriguing that these two sectors with conserved charges are precisely the seeds that generate all the multipoles with vanishing linear response via the ladder structure. In fact, by exploiting the existence of the ladders, we can construct two series of conserved charges that connect with the monopole and the dipole. Let us see each one separately.

\subsubsection{Monopole}
Starting from the monopole, we can use the big ladder to go straight up to the tower of multipoles $\ell=2(d-1)n$ with $n$ a strictly positive integer. On the other hand, we can connect it to the multipole $\ell=d$ via the small ladder and, from there, again all the way up the tower with $\ell=2(d-1)n+d$. Let us see how the construction works more explicitly. The general solution for the monopole can be written as
\be
\Phit_0=\Qb_0 (-z)^{\frac{1}{2(d-1)}}-2(d-1) Q_0\;_2F_1\left(-\frac{1}{2},-\frac{1}{2(d-1)},1-\frac{1}{2(d-1)};z\right)\,,
\ee
where $Q_0$ and $\Qb_0$ are the constants of integration which have been chosen so that $Q_0$ is the monopole conserved charge, i.e. $\mathcal{Q}_0[\Phit_0]=Q_0$, and $\Qb_0$ is the mode with vanishing charge. We can now easily identify that the regular solution corresponds to the vanishing charge solution $Q_0=0$ so that we conclude that regularity selects the sector with a trivial charge. 

We can now act with the ladders to generate solutions for other multipoles. By applying the small ladder operator $a_{d,0}^+$ we obtain a solution for the monopole $\ell=d$. Since the ladder preserves the regularity of the solutions, the regular solution of the monopole will be connected to the regular solution of $\ell=d$. From there, we can then apply the big ladder $A_{d,\ell=d}^+$ to generate the corresponding tower of regular solutions that are connected to the monopole. Furthermore, since the seed monopole solution has a vanishing charge, all the regular solutions of this tower will be assigned a vanishing charge as well. Let us notice that this tower of solutions are precisely the ones with vanishing linear response. We can understand this in terms of the ladders and the relation to the monopole charge. Since the regular monopole solution with vanishing charge is simply a power of $z$, when applying the ladder operators we will always generate solutions that are polynomials (up to some fractional power of $z$) and this prevents the appearance of a decaying tail that would describe the linear response and, consequently, we have $\alpha_\ell=0$ for all these multipoles.

Alternatively, we can act with the big ladder directly on the monopole solution to generate another tower of solutions. However, we encounter the obstruction already discussed above. If we apply $A_{d,0}^+$ to the monopole solution, we obtain
\be
A^+_{d,0}\Phit_0=Q_0(1-z)^{3/2}\,,
\ee
so we only generate half of the space of solutions of the multipole $\ell=2(d-1)$. This is again due to the fact that the mode $\Qb_0$ turns out to be in the kernel of $A^+_{d,0}$ so it is projected out. Since $\Qb_0$ is the regular mode, we are projecting out the physical solution which, then, might appear as disconnected from the monopole solution. However, we can climb down instead of up. If we start from the general solution for $\ell=2(d-1)$ written as
\be
\Phit_{\ell=2(d-1)}=Q_0 (1-z)^{\frac{3}{2}}+\Qb_{2(d-1)}(-z)^{\frac{1}{2(d-1)}}\;_2F_1\left(1,\frac{3d-4}{2(d-1)},1+\frac{1}{2(d-1)};z\right)\,,
\ee
and apply $A^-_{d,0}$ to descend to $\ell=0$, we project out the singular mode and, thus, we descend to the regular monopole solution. Therefore, although we cannot climb up from the regular monopole solution to the regular solution of $\ell=2(d-1)$, we can connect both regular sectors by means of $A^-_{d,\ell=2(d-1)}$. In other words, although we have $A^+_{d,0}\Phit_0^{\text{reg}}=0$, the relation $A^-_{d,0}\Phit_{2(d-1)}^{\text{reg}}=\Phit_{0}^{\text{reg}}$. This shows that the connection between both sectors is not both ways, but only downstairs. Analogously, we can start from an arbitrary multipole in that tower and the regular solution will then be eventually connected to the monopole regular sector by going downwards with the corresponding lowering operators. In this case, the regular solutions do not exhibit a polynomial behaviour and this explains why this tower of solutions do not have vanishing linear response.

\subsubsection{Dipole}
The dipole potentially provides a seed to generate other two towers of solutions analogous to the monopole. We can use the big ladder directly to generate all the multipoles with $\ell=2(d-1)n+1$ or employ first the small ladder to connect with $\ell=d-1$ and from there climb up with the big ladder to generate $\ell=2(d-1)n+d-1$. In terms of the dipolar conserved charge $Q_1$, the solution can be written as
\be
\Phit_1=\Qb_1+2(d-1)Q_1(-z)^\frac{1}{2(d-1)}\;_2F_1\left(-\frac{1}{2},\frac{1}{2(d-1)},1+\frac{1}{2(d-1)};z\right)\,
\ee
where $\Qb_1$ is the solution with vanishing charge and we have $\mathcal{Q}_1[\Phit_1]=Q_1$. In this case, the regular solution corresponds to $\Qb_1=0$ so it is given by the sector with a non-vanishing conserved charge $Q_1\neq0$.

The action of the big ladder on the full solution yields
\be
A_1^+\Phit_1=Q_1(1-z)^{3/2}(-z)^\frac{1}{2(d-1)}
\ee
so it annihilates the singular sector and we obtain $A_1^+\Phit_1=\Phit_{2d-1}^{\text{reg}}$. Thus, in this case we can climb up the big ladder and generate all the regular solutions of this tower by the iterative application of the big ladder raising operators. The obstruction comes this time when trying to climb down the ladder. If we consider the general solution $\Phit_{2d-1}$, we have that $A_{d,1}^-\Phit_{2d-1}=\Phit_1^{\text{sing}}$, i.e., it projects out the regular solution and, consequently, we cannot take the regular solution of $\ell=2d-1$ down to the ground floor $\ell=1$. Let us also notice that the regular solution obtained for $\ell=2d-1$ is again a sum of powers of $z$ and the application of the raising operators will maintain this property. The consequence of this is that all the modes in this tower will not have decaying tails, thus explaining the vanishing of the linear response for the multipoles $\ell=2(d-1)n+1$ with $n=1,2,3,\cdots$. Furthermore, since the seed of this tower, the dipole, does not share this property, it does have a linear response, again in agreement with our direct calculation.

Let us now turn to the multipoles connected via the small ladder. This ladder allows to connect to $\ell=d-1$ and from there climb up again with the big ladder. These connections do not present any obstructions and the regular solutions of the dipole seed the regular solutions of the subsequent multipoles. Furthermore, the generated multipoles do not acquire the form of a sum of power laws, so the linear response does not vanish in general.

\section{On the existence of ladders}
\label{Sec:existence}

In the previous sections we have unveiled and analysed the  ladder operator structure and their symmetries in the DBI theory. A natural question to ask is how general our result is and if other types of ladder structures are possible. It could be that the particular two-ladder structure that we have obtained is related to some special choice of coordinates or field normalisation. In this section, we will explore in a more systematic manner the existence of ladder operators. For that,
we will return to the equation for the multipoles that we will write as
\be
M(x)\frac{\dd^2 \Phi_\ell}{\dd x^2}+N(x)\frac{\dd \Phi_\ell}{\dd x}-\ell(\ell+d-2)\Phi_\ell=0\,,
\label{Eq:GeneralEquation}
\ee
where
\begin{align}
&M(x)=x^2 c_r^2(x)=x^2\Big(1+x^{2(1-d)}\Big),\\
&N(x)=-x^2 c_r^2 \partial_x\log\Big(x^{d-3} \mK_X\Big)=x\Big(3-d+2x^{2(1-d)}\Big).
\end{align}
In the following, we will keep the functions $M(x)$ and $N(x)$ general to make apparent where the special properties of the DBI theory enter. We will encounter some miracles that occur for DBI and we will point out where they appear.

The general idea now will be to bring Eq.\eqref{Eq:GeneralEquation} to a canonical form. For that, we will introduce a variable $z=z(x)$ and a transformation $\Phi_\ell(x)\to U(x)\Phi_\ell(x)$ so the equation reads
\begin{align}
&M(x) z'^2(x)\frac{\dd^2 \Phi_\ell}{\dd z^ 2}+\left[N(x)z'(x)+M(x) z''(x)+2M(x) z'^2(x)\frac{U_z}{U}\right]\frac{\dd \Phi_\ell}{\dd z}\nonumber\\
&-\left[\ell(\ell+d-2)-M(x) z'^2(x)\frac{U_{zz}}{U}-\Big(N(x) z'(x)+M(x) z''(x)\Big)\frac{U_z}{U}\right]\Phi_\ell=0.
\end{align}
We want to bring this equation to the form of a hypergeometric equation. For that, we will choose the new variable to satisfy:
\be
M(x) z'^2= \lambda z(z-1),
\label{eq:eqM}
\ee
where $\lambda$ is some constant parameter. This parameter could depend on the multipole $\ell$, but we will take it to be the same for all multipoles. The domain of the variable $z$ is determined by the sign of $\lambda$ because $M(x)$ is a positive function (provided we exclude Laplacian instabilities). Thus, for positive $\lambda$  we have that either $z\geq1$ or $z\leq0$. On the other hand, if $\lambda$ is negative then $z\in [0,1]$.  In fact, these two branches are connected via the transformation $z\to 1-z$ and $\lambda\to-\lambda$ so without loss of generality we can focus on $z>0$.

We can integrate Eq. \eqref{eq:eqM} for an arbitrary function $M(x)$ so the solution can be expressed as
\be
\int\frac{\dd z}{\sqrt{\lambda z(z-1)}}=\pm\int\frac{\dd x}{\sqrt{M(x)}}.
\label{Eq:GensolM}
\ee
For the DBI case, we can perform the change of variables $y=a+b x^{2(d-1)}$ to obtain
\be
\int\frac{\dd z}{\sqrt{\lambda z(z-1)}}=\pm\int\frac{\dd x}{\sqrt{x^2\big(1+x^{2(1-d)}}\big)}=\pm\int\frac{\dd y}{2(d-1)\sqrt{(y-a)(y+b-a}\big)}.\nonumber
\ee
so we can identify $z$ and $y$ if we pick the positive branch, $\lambda=4(d-1)^2$ and either $(a=0,b=-1)$ or $a=b=1$ that correspond to the following two changes of variables:
\be
z=1+x^{2(d-1)}\quad z=-x^{2(d-1)}.
\ee
The second change of variables is the one we used in the previous sections and it corresponds to the branch with $z<0$. The first one corresponds to the branch with $z>1$. Notice that they are related by $z\to 1-z$, which is a symmetry of the hypergeometric equation, so we will focus on the first change of variables from now on. Thus, we will have that $z\geq1$ so the origin will be placed at $z=1$. For this change of variables, it is straightforward to verify that the DBI theory satisfies
\be
Nz'=2(d-1)\Big[d-1-(d-3)z\Big]\,\qquad Mz''=2(d-1)(2d-3) z.
\ee
Here we find one of the first little miracles of DBI consisting in the fact that the change of coordinates is a simple power law transformation and the combinations $Nz'$ and $Mz''$ become linear polynomials. For this to happen, some conspiracies occurred. The transformed equation in the transformed variable then becomes
\begin{align}
&4(d-1)^2 z (z-1)\frac{\dd^2 \Phi_\ell}{\dd z^ 2}+2(d-1)^2\left[1+\frac{d}{d-1}z+4z(z-1)\frac{U_z}{U}\right]\frac{\dd \Phi_\ell}{\dd x}\nonumber\\
&-\left[\ell(\ell+d-2)-4(d-1)^2z(z-1)\frac{U_{zz}}{U}-2(d-1)^2\left(1+\frac{d}{d-1}z\right)\frac{U_z}{U}\right]\Phi_\ell=0.
\end{align}
To recover the form of an hypergeometric equation, the coefficient
of $\dd\Phi_\ell/\dd z$ has to be a linear polynomial so we need to impose
\be
2(d-1)^2+2(d-1)(2d-3)z+8(d-1)^2z(z-1)\frac{U_z}{U}=N_0+N_1 z
\ee
for some constants $N_0$ and $N_1$. This equation can be integrated to obtain
\be
U(z)=U_0 z^m(z-1)^n,
\ee
with
\be
m=\frac14 - \frac{N_0}{8 (d-1)^2},\qquad n=\frac{-2 + 6 d - 4 d^2 + N_0 + N_1}{8 ( d-1)^2}
\ee
We can replace this solution into the transformed equation to obtain
\begin{align}
    &4(d-1)^2 z (z-1)\frac{\dd^2 \Phi_\ell}{\dd z^ 2}+\Big(N_0+N_1 z\Big)\frac{\dd \Phi_\ell}{\dd x}
    &-\Big[\ell(\ell+d-2)+\frac{m_0+m_1z+m_2z^2}{16 (d-1)^2 z(z-1)}\Big]\Phi_\ell=0,
\end{align}
with
\begin{align}
&m_0=(2 - 4 d + 2 d^2 - N_0) (10 - 20 d + 10 d^2 + N_0),\\
&m_1=-2 (16 - 60 d + 84 d^2 - 52 d^3 + 12 d^4 - 8 N_0 + 16 d N_0 - 8 d^2 N_0 +
    N_0 N_1),\\
&m_2=-(-2 d + 2 d^2 - N_1) (8 - 14 d + 6 d^2 - N_1).
\end{align}
We then need to fix the constants $N_0$ and $N_1$ so the coefficient of $\Phi_\ell$ does not depend on $z$. We find the four solutions
\begin{align}
&\text{A:}\qquad N_0= -10(d-1)^2,\quad N_1=2(d-1)(7d-8),\nonumber\\
&\text{B:}\qquad N_0= -10(d-1)^2,\quad N_1=2(d-1)(7d-6),\nonumber\\
&\text{C:}\qquad N_0= 2(d-1)^2,\qquad N_1=2(d-1)(d-2),\nonumber\\
&\text{D:}\qquad N_0= 2(d-1)^2,\qquad N_1=2d(d-1).\nonumber
\end{align}
Thus, we finally obtain the hypergeometric equation
\begin{align}
    4(d-1)^2 z (z-1)\frac{\dd^2 \Phi_\ell}{\dd z^ 2}+\Big(N_0+N_1 z\Big)\frac{\dd \Phi_\ell}{\dd x}
    -\Big[\ell(\ell+d-2)+\alpha_a\Big]\Phi_\ell=0,
\end{align}
with 
\be
\alpha_\text{A}=2(1-d)(3d-4),\quad \alpha_\text{B}=3(1-d)(2d-1),\quad\alpha_\text{C}=1-d,\quad\alpha_\text{D}=0,
\label{eq:alphaDBI}
\ee
for the four found branches. We can then proceed to obtain the ladder operators by introducing the associated family of Hamiltonians
\be
H_{\ell}\equiv -M(z)\left[M(z)\frac{\dd^2}{\dd z^2}+N(z)\frac{\dd }{\dd z}-\ell(\ell+d-2)-\alpha\right],
\ee
with
\be
M(z)=\lambda z(z-1),\quad N(z)=N_0+N_1 z\,
\label{Eq:defMN}
\ee
and the values for $\alpha$ obtained in \eqref{eq:alphaDBI}. We will keep all the parameters general and will only specialise to the DBI case at the end in order to illustrate again how DBI stands out as a very special case. We will follow the same procedure as in \cite{Jimenez:2022mih} that consists in factorising the Hamiltonian in terms of ladder operators as
\bea
\Am_\ell \Ap_\ell&=&H_\ell+\varepsilon_{1\ell},\nonumber\\
\Ap_\ell \Am_{\ell}&=&H_{\ell+n}+\varepsilon_{2\ell},
\label{Eq:FactorizationH}
\eea
with $\varepsilon_{i,\ell}$ some scalars and $n$ is some integer number that will determine the separation between the multipoles connected by the ladder operators, i.e., the size of the ladder steps. To obtain the factorisation \eqref{Eq:FactorizationH}, we make the Ansatz
\bea
\Am_\ell&\equiv& M(z)\partial_z+W_{1,\ell}(z),\\
\Ap_\ell&\equiv&-M(z)\partial_z+W_{2,\ell}(z),
\eea
where the functions $W_{i,\ell}$ are to be determined by requiring the desired factorisation. We then find
\bea
\Am_\ell \Ap_\ell&=&H_\ell+M\Big(W_{2,\ell}-W_{1,\ell}+N-M'\Big)\partial_z\nonumber\\
&&+\left[W_{1,\ell}W_{2,\ell}-M\Big(\ell(\ell+d-2)+\alpha-W'_{2,\ell}\Big)\right],\label{Eq:AmAp}\\
\Ap_\ell \Am_\ell&=&H_{\ell+n}+M\Big(W_{2,\ell}-W_{1,\ell}+N-M'\Big)\partial_z\nonumber\\
&&+\left[W_{1,\ell}W_{2,\ell}-M\Big((\ell+n)(\ell+d-2)+\alpha+W'_{1,\ell}\Big)\right].\label{Eq:ApAm}
\eea
The vanishing of the coefficient of $\partial_z$ requires
\be
W_{2,\ell}-W_{1,\ell}=M'-N.
\label{Eq:W2W1}
\ee
If we subtract \eqref{Eq:AmAp} and \eqref{Eq:ApAm} and use the relation \eqref{Eq:W2W1} we find
\be
M\Big[2W'_{1,\ell}-N'+M''+n(n+2\ell+d-2)\Big]=\varepsilon_{1,\ell}-\varepsilon_{2,\ell}.
\ee
As argued in \cite{Jimenez:2022mih}, the solution for $W_{1,\ell}$ is regular if we impose $\varepsilon_{1,\ell}=\varepsilon_{2,\ell}\equiv\varepsilon_\ell$ so the RHS vanishes\footnote{Otherwise there is a particular solution that goes as $(\varepsilon_{1,\ell}-\varepsilon_{2,\ell})\arctan(1-2z)$ that jeopardizes the good behaviour of $W_{1,\ell}$.} and we obtain
\be
W_{1,\ell}=c_{\ell}-\frac{n}{2}(n+2\ell+d-2)+\frac{N-M'}{2},
\ee
with $c_\ell$ some constant of integration. We then find
\bea
\varepsilon_\ell&=&\frac{n}{4}\big(n+2\ell+d-2\big)\Big[n\big(n+2\ell+d-2\big)z-4c_\ell\Big]z\nonumber\\
&&+\frac14\left[4c_\ell^2-\big(N-M'\big)^2-2M\Big(2\ell^2+(2\ell+n)(n+d-2)+2\alpha+N'-M''\Big)\right].
\eea
So far, we have not used the explicit form of the functions $M$ and $N$. We can now use their explicit forms \eqref{Eq:defMN} to obtain
\bea
\varepsilon_\ell&=&\varepsilon^{(0)}_\ell+\varepsilon^{(1)}_\ell z+\varepsilon^{(2)}_\ell z^2,
\eea
with
\bea
\varepsilon^{(0)}_\ell&=&c_\ell^2-\frac14(N_0+\lambda)^2\\
\varepsilon^{(1)}_\ell&=&-\frac{N_0 N_1}{2} + \ell^2 \lambda + 
 \ell (n+d-2) \lambda + (N_0 + \alpha) \lambda + 
 \frac{n}{2} (n+d-2) (\lambda - 2 c_\ell) - 2 \ell n c_\ell\nonumber\\
\varepsilon^{(2)}_\ell&=&\frac14n^4+\frac12(2\ell+d-2\ell)n^3+\frac14\left[\big(2\ell+d-2\big)^2-2\lambda\right]n^2-\frac\lambda2\big(2\ell+d-2)n\nonumber\\
&+&\frac14N_1\big(2\lambda-N_1\big)-\lambda\big[\alpha+\ell(\ell+d-2)\big].\nonumber
\eea
Since $\varepsilon_\ell$ cannot depend on $z$, we need to impose $\varepsilon^{(1)}=0$, that fixes the constant $c_\ell$, and $\varepsilon^{(2)}=0$, which can be seen as an equation for $n$ that fixes the size of the ladder. We then find that the existence of ladder operators need to jump in $\ell-$steps of size $n$ with
\be
n=\frac{1}{2}\left[2-d-2\ell\pm\sqrt{\big(2\ell+d-2\big)^2+4\lambda\pm4\sqrt{4\lambda\ell^2+4(d-2)\lambda\ell+N_1(N_1-2\lambda)+\lambda(\lambda+4\alpha)}}\right].
\ee
In order to have a proper ladder structure, we need $n$ to be some positive integer. In order to guarantee that we will impose the radicands to be perfect squares. Since we are assuming that all the parameters are independent of $\ell$, we find that the inner radical must take the form
\be
4\lambda\ell^2+4(d-2)\lambda\ell+N_1(N_1-2\lambda)+\lambda(\lambda+4\alpha)=\lambda\Big(2\ell+d-2\Big)^2
\ee
that requires to have
\be
(d-2)^2\lambda=N_1^2-2N_1\lambda+\lambda(\lambda+4\alpha).
\label{eq:ladderconstraint1}
\ee
This condition imposes a non-trivial constraint on the form of the equation in order to have the desired ladder structure. In principle, this can be used to determine which $K-$essence theories admit such a structure and it is straightforward to check that the values of $\lambda$, $N_1$ and $\alpha$ obtained above for DBI indeed fulfill \eqref{eq:ladderconstraint1}. Furthermore, upon use of the relation \eqref{eq:ladderconstraint1},  the outer radical also becomes a perfect square and we finally find the following expression for the size of the steps:
\be
n=\frac{1}{2}\left[2-d-2\ell\pm\Big(2\ell+d-2\pm2\sqrt{\lambda}\Big)\right].
\ee
This expression shows that the size of the steps only depends on the parameter $\lambda$ which in turn needs to  be a perfect square. For the DBI theory this is indeed satisfied since we have $\lambda=4 (d-1)^2$. For this value of $\lambda$, we find the four possibilities
\begin{align}
n_{++}=2(d-1),\quad n_{+-}=2(1-d),\quad n_{-+}=4-3d-2\ell,\quad n_{--}=d-2\ell.
\end{align}
The first two cases reproduce the raising and lowering operators of the big ladder, while the fourth case gives the small ladder. The third solution is not physical for $d>2$ since the connection would be to a multipole with $\ell<0$. We have thus reproduced the whole ladder structure of the DBI. We have however shown that these systems can admit generic ladder structures depending on the value of $\lambda$. In general, we find the four branches of solutions
\be
n_{+\pm}=\pm\sqrt{\lambda},\quad n_{-\pm}=2-d-2\ell\mp\sqrt{\lambda}.
\ee
It is clear that the two branches describe a big ladder structure where the size of the step is fixed and determined by $\lambda$, which needs to be a perfect square, but we can, in principle, construct a ladder of any size by choosing $\lambda$. On the other hand, the other branches of solutions represent a small ladder because the size depends on $\ell$. Again, the branch $n_{-+}$ is not physical because it will involve a connection to negative values of $\ell$. The remaining branch connects $\ell$ to $(2+n)-d-\ell$ with $n=\sqrt{\lambda}$ a non-negative integer, so we see that it generalises the small ladder structure. Let us notice that $n$ could depend on the dimension $d$ as it happens in the DBI theory. In any case, it is clear that these systems admit a rich ladder structure that would be interesting to explore further. In particular, we could ask the question of what ladder structures can be obtained from a given physical system or, in other words, if given a ladder structure we can find a physical system described by it.

\section{Discussion and conclusions}
\label{Sec:Discussion}

This work has been devoted to performing a detailed analysis of static perturbations around screened point-like objects for the DBI scalar field theory in arbitrary dimension. This theory has been shown to be exceptional among theories featuring a $K$-mouflage screening and our results on different aspects of the static perturbations adds to this set of remarkable properties. One of these has been found for the linear response where we have obtained two sets of special multipoles, namely: multipoles with vanishing linear response and multipoles with a logarithmic running. The set of multipoles with the said properties form a regular pattern in the parameter space $(\ell,d)$. We have found that the deformability coefficients vanish for some multipoles in any space dimension $d$, although the multipoles with vanishing deformability are scarcer as  the dimension increases, having its most complete structure in $d=3$ where all odd multipoles above the dipole have vanishing deformability. On the other hand, multipoles with a logarithmic running only exist in even dimension, although they share the property of being scarcer as the number of dimensions increases. The existence of multipoles with vanishing linear response can be interpreted as a consequence of imposing a rigid boundary condition at the position of the particle. A singular property of DBI is the improved regular behaviour of the field profile at the position of the particle and this property has motivated us to impose regular boundary conditions for the perturbations at the particle position. These boundary conditions precisely select the regular solutions that lack a decaying tail in the asymptotic region for certain multipoles, thus causing the vanishing of the linear response. In this respect, it is noteworthy that the requirement of regularity at the horizon of BHs is what selects the solutions without an asymptotically decaying tail and, hence, gives rise to the vanishing of certain Love numbers. Thus, there is a clear analogy of the physical properties of both systems that calls for a deeper exploration.

We have shown that the equations for the multipoles can be conveniently expressed via a family of Hamiltonians that admit a factorisation in terms of ladder operators. Although this by itself can be regarded as an interesting feature, what is more interesting is how the ladder operators are organised. Remarkably, we have found that the DBI system admits a ladder structure formed by two sets of ladder operators in any dimension $d$. This structure is the generalisation to arbitrary dimension of the analogous structure unveiled for its spin-1 Born-Infeld  relative in $d=3$ \cite{Jimenez:2022mih}. An intriguing property of the big ladder structure is that the the lowest multipole (monopole and dipole) ladder operators have kernels with a non-trivial overlap with the corresponding Hamiltonians so that they project out certain sectors of the space of solutions. Another interesting property of the ladder structure is that the series of lacunar multipoles with vanishing or logarithmically running linear response can be connected via the ladders. This is thanks to the property of the ladder operators of preserving the regularity and the polynomiality of the solutions for the multipoles. Furthermore, we have shown the existence of conserved charges for the monopole and the dipole which, via the ladder structure, can be extended to charges for all the multipoles. These conserved charges give yet another view on the special properties of the linear response because they allow to connect the behaviour of the solutions near the particle (where boundary conditions are set) and their behaviour in the asymptotic region (where the linear response is defined). One pertinent question at this point is to what extent the ladder structure is special and whether other structures could be possible. We have answered this question by considering a fairly general family of Hamiltonians and carrying out a general procedure to obtain a factorisation in terms of ladder operators. The DBI theory has shown its exceptional character by giving rise to some crucial occurrences in our derivation that permits to construct ladder structures. Our results have shown that the two-ladder structure of DBI is unique. However, our procedure has also revealed the potential existence of systems where the two-ladder structure could differ from the DBI one. It remains to corroborate if those structures could correspond to physical systems, but we leave this for future work.

The techniques that we have presented here could be extended to other physically relevant situations such as e.g. when the background geometry is not Minkowski. A particular scenario that we have in mind is how our results are modified when the point-like object is replaced by an object with a gravitational horizon, i.e., black holes or de Sitter space-time. It will be interesting to investigate if and to what extent the singular properties that we have found persist for more general backgrounds. Likewise, it would be interesting to analyse if the vanishing of the linear response for certain multipoles also persist at higher orders as is the case for Schwarzschild black holes \cite{Riva:2023rcm}. It is known that DBI theory is special because it possesses an enhanced symmetry (see e.g. \cite{Pajer:2018egx}) and we could conjecture that this symmetry might be behind the vanishing of the linear response and could guarantee the persistence of this property at full non-linear order. If the enhanced symmetry is really relevant, other special shift-symmetric scalar theories with enhanced symmetries as those obtained in \cite{Pajer:2018egx,Grall:2019qof} should be investigated. On the other hand, these symmetries should be realised at the level of the effective field theory capturing the coupling between the screened objects and external fields, i.e, the vanishing of certain couplings for certain multipoles should arise from the action of the symmetries on the external fields. We have not touched upon this subject and would like to come back to these issues in the future. 

Many of the results obtained in this work for DBI are shared by its spin-1 counterpart, namely Born-Infeld non-linear electromagnetism (see \cite{Jimenez:2022mih}). We have explained the close relation between both theories in Sec. \ref{Sec:relationtoBI} and they share other remarkable properties such as maintaining causal propagation (see e.g. \cite{Deser:1998wv,Gibbons:2000xe,Mukohyama:2016ipl,deRham:2016ged}). In fact, this property can be used to select essentially Born-Infeld and DBI so this could ultimately be the cause for the exceptional behaviour found for these theories, although an explicit link is still missing. In this respect, we would find interesting to analyse how our results would change for more general gravitational backgrounds for Born-Infeld electromagnetism. In particular, the coupling between Einstein gravity and Born-Infeld electrodynamics should reveal extended ladder symmetries which could combine the cases of black holes and screening of the electromagnetic interaction. The former is characterised by a typical length scale, e.g. the Schwarzschild radius, and the latter by the screening radius around a point particle charge. Combining both effects and studying the deformation of this geometry vs the electromagnetic field distribution is something we would like to analyse. The case of rotating black-holes coupled to non-linear electrodynamics would also be of interest and may have astrophysical applications. Finally, the underlying geometry corresponding to the ladder structures and their physical consequence for the long wavelength physics around screened objects is also worth investigating~\cite{BenAchour:2022uqo}.

\section*{Acknowledgments}

We thank Luca Santoni for useful discussions. JBJ and DB acknowledge support from Project PID2021-122938NB-I00 funded by the Spanish “Ministerio de Ciencia e Innovaci\'on” and FEDER “A way of making Europe”. DB acknowledge support from PIC-2022-02 funded by Salamanca University and PID2022-139841NB-I00 funded by the Spanish  “Ministerio de Ciencia e Innovaci\'on". 

\appendix

 \section{K-mouflage and supersymmetric quantum mechanics}
\label{Appedix}
 
The results of the main text can be obtained using an alternative method where supersymmetry is emphasized ~\cite{Cooper:1994eh,gangopadhyaya2011supersymmetric}.  
  \subsection{The supersymmetric Lagrangian}

Perturbing around the background solution $\bar\phi$ as 
$ 
\phi= \bar\phi +\varphi
$
we obtain the action at second order for each mode
\be 
S= S_{d-1}\int \dd r \dd t \left( \frac{1}{2}\frac{\mK_X}{c^2}(\partial_t \Phi
)^2 - \frac{1}{2}(\partial_r \Phi)^2 - \frac{\tilde M^2 }{2}\Phi^2\right)
\ee
where $\Phi= r^{(d-1)/2} c\varphi $ is dimensionless and we have decomposed the fields in Legendre polynomials. We have introduced
\be 
\tilde M^2 = \partial^2_r \ln r^{(d-1)/2 c}+ (\partial_r \ln r^{(d-1)/2} c)^2 + \mK_X\frac{\ell(\ell+d-2)}{r^2 c^2}
\ee
and the speed of perturbations
$
c^2 = \mK_X + 2X \mK_{XX}
$
evaluated at the background level. Here $S_{d-1}$ is the volume of the $(d-1)$-sphere. 
Let us specialise to the static case and define the rapidity variable
$ 
\frac{dz}{dr}= \frac{\sqrt \mK_X}{r c}
$
and the new field
\be 
\Psi= \left(\frac{dz}{dr}\right)^{1/2}\Phi\,,
\ee
whose action becomes
\be 
S= -S_{d-1}\int dt dz \left( \frac{1}{2} (\partial_r \Psi)^2 + \frac{m^2}{2} \Psi\right)\,,
\ee
where the mass is now
$
m^2= \ell(\ell+d-2) + W^2 + \partial_z W\,.
$
The superpotential $W$ is given by
\be 
W= \frac{1}{2}\partial_z \ln \frac{c\sqrt{\mK_X}}{ r^{2-d}}\,.
\ee
The mode equation of motion can be written as an eigenvalue problem
\be 
H\Psi= - \ell(\ell+d-2) \Psi
\ee
where the Hamiltonian is the one of supersymmetric quantum mechanics with
\be 
H= Q^\dagger Q
\ee
where
\be 
Q= -\frac{d}{dz}+W.
\ee
Notice that $H$ is positive definite operator in the Hilbert space $L^2$. As we are looking for negative eigenvectors, they cannot belong to the Hilbert space and therefore we will find unbounded solutions corresponding to the excitation of non vanishing scalar fields at infinity. 

\subsection{DBI and Natanzon potentials}
We now consider the DBI case for which
\be 
\mK= \sqrt{1+2X}\,.
\ee
We focus on the case of a point particle source such that the background solution becomes solution to
\be 
\mK_X \partial_r \phi= \frac{\beta M}{S_{d-1}M_{\rm Pl}^{(d-1)/2}} \frac{1}{r^{d-1}}
\ee
In the DBI case, and defining the K-mouflage radius
\be 
r_s^{d-1}= \frac{\beta M}{S_{d-1}\Lambda^{(d+1)/2}M_{\rm Pl}^{(d-1)/2}}
\ee
we have 
\be 
X= -\frac{1}{2}\frac{1}{1+ (\frac{r}{r_s})^{2(d-1)}}
\ee
Defining
\be 
\sinh \theta =\left(\frac{r}{r_K}\right)^{(d-1)}
\ee
we find that the rapidity variable is
\be 
z= \frac{\theta}{d-1}
\ee
and the superpotential becomes
\begin{equation}
    W = -\frac{d}{2t}+(d-1) t
\end{equation}
where
\be 
t= \tanh (d-1) z.
\ee
Notice that this potential agrees with that in Eq.~\eqref{eq:PT_potential} once the coefficients are rescaled by a factor $1/(d-1)$.
\subsection{The spectrum of DBI perturbations as supersymmetric states}
Let us notice that the superpotential $W$ belong to the general family of superpotentials
\be 
{\cal W}_{a,b}= \frac{a}{t} + b t\,,\qquad t=\tanh \kappa z\,.
\ee
The DBI case is obtained for
\be 
a= -\frac{d}{2},\ b= d-1, \kappa= d-1.
\ee
We define the  ladder operators 
\be 
Q_{a,b} = -\frac{d}{dz}- {\cal W}_{a,b}\,,\quad  Q^\dagger_{a,b}= \frac{d}{dz}- {\cal W}_{a,b}\,,
\ee 
such that
\be \label{eq:HamNat}
{\cal H}_{a,b}\equiv Q^\dagger_{a,b}Q_{a,b}=-\frac{d^2}{dz^2}+ {\cal W}_{a,b}^2 -\frac{d{\cal W}_{ab}}{dz}=-\frac{d^2}{dz^2}+ V_{a,b}\,,
\ee
where we have introduced the potential $V_{a,b}$ that has the explicit form
\be 
V_{a,b}=  (a+b)^2 + \frac{a(a+\kappa)}{s^2}-\frac{b(b+\kappa)}{c^2}\,,
\ee
where $c=\cosh \kappa z$ and $s=\sinh \kappa z$. There is an interesting set of symmetries enjoyed by this potential. These symmetries correspond to changing $a\to -a-\kappa$ and $b\to -b-\kappa$. These transformations only change the potential by a constant and we can obtain the following families of related potentials:
\bea
V_{a,b}&=&V_{-a-\kappa,b}+(2b-\kappa)(2a+\kappa),\\
V_{a,b}&=&V_{a,-b-\kappa}+(2a-\kappa)(2b+\kappa),\\
V_{a,b}&=&V_{-a-\kappa,-b-\kappa}-4\kappa(a+b+\kappa).
\eea
which allows to obtain eigenvectors of the Hamiltonian defined by $V_{a,b}$ from eigenvectors of the related potentials with the corresponding substitutions. In the following we denote by
$
\tilde a= -\kappa -a, \ \tilde b= -\kappa -b
$
and we obtain the new eigenvalues $\tilde c(a,b)$ from the eigenvalues $c(a,b)$ of ${\cal H}_{a,b}$. We then find the following families of eigenvalues:
\bea
c_{\rm I}(a,b)&=&c(a,b),\\
c_{\rm II}(a,b)&=& c(\tilde a, b) + (2b-\kappa)(2a+\kappa)\,,\\
 c_{\rm III}(a,b)&=& c(\tilde a,\tilde b) - 4\kappa(\kappa+a+b)\,,\\
c_{\rm IV}(a,b)&=& c(a,\tilde b)+ (2a-\kappa)(2b+\kappa)\,.
\eea
This constructs four sets of eigenvalues and eigenvectors for ${\cal H}_{a,b}$.

For  arbitrary values of $a$ and $b$, we will find  eigenvalues and eigenvectors for bound states of the Hamiltonian ${\cal H}_{a,b}$. From Eq. \eqref{eq:HamNat}
we obtain  the explicit ladder identity
\be 
Q_{a,b}Q_{a,b}^\dagger= 4\kappa(a+b-\kappa) + {\cal H}_{a-\kappa,b-\kappa}\,.
\label{recur1}
\ee
Notice that the action of $Q_{a,b}Q^\dagger_{a,b}$ lowers the parameters of the Hamiltonian by $\kappa$. 

As usual in supersymmetric systems, we introduce the vacuum state as being in the kernel of the supersymmetric operator $Q_{\alpha,\beta}$ for a given choice of the  indices $(\alpha,\beta)$. Here we introduce the vacuum state by the property
\be 
Q_{a-n\kappa,b-n\kappa}\vert f_0^{\rm I}\rangle =0\,,
\ee
where $n$ is an  integer which is not specified yet. 
Explicitly the  wave function reads
$ 
f_0^{\rm I}(z)= e^{-\int dx {\cal W}_{a-n\kappa,b-n\kappa}(z)}\,.
$
We can also introduce the 
excited states using the ladder operators
\be 
\vert f_n^{\rm I}\rangle= Q^\dagger_{a,b} Q^\dagger_{a-\kappa,b-\kappa}\dots Q^\dagger_{a -(n-1)\kappa,b-(n-1)\kappa}\vert f_0^{\rm I}\rangle\,.
\ee
Using the recursion relation (\ref{recur1}) we find that this excited state is an eigenstate of ${\cal H}_{a,b}$, i.e.  we have
\be 
{\cal H}_{a,b}\vert f_n^{\rm I}\rangle  = c_n(a,b) \vert f_n^{\rm I}\rangle\,,
\ee
with 
\be 
c_n(a,b)= 4n\kappa (a+b-\kappa n)\,.
\ee
Finally notice that  the wave function  is given explicitly by
\be 
f_n^{\rm I}(z)= (\vert \sinh \kappa z\vert )^{-(a-n\kappa)/\kappa}
( \cosh \kappa z)^{-(b-n\kappa)/\kappa}\,,
\ee
which is an even function. As the ladder operators are odd, the excited states are either odd or even depending on the parity of $n$.

We can now then construct three more series of eigenvectors and eigenvalues defined by
\bea 
Q_{\tilde a-n\kappa, b-n\kappa}\vert  f_0^{\rm II}\rangle &=&0\,,\\
Q_{\tilde a-n\kappa,\tilde b-n\kappa}\vert f_0^{\rm III}\rangle &=&0\,,\\
Q_{ a-n\kappa,\tilde b-n\kappa}\vert  f_0^{\rm IV}\rangle &=&0\,,
\eea
where $n$ is an  integer which is not specified yet. Analogously, we can also introduce the
excited states using the ladder operators constructed out of the tilded quantities
\bea
\vert  f_n^{\rm II} \rangle&=& Q^\dagger_{\tilde a, b} Q^\dagger_{\tilde a-\kappa, b-\kappa}\dots Q^\dagger_{\tilde a -(n-1)\kappa, b-(n-1)\kappa}\vert  f_0^{\rm II}\rangle\,,\\
\vert  f_n^{\rm III}\rangle &=& Q^\dagger_{\tilde a,\tilde b} Q^\dagger_{\tilde a-\kappa,\tilde b-\kappa}\dots Q^\dagger_{\tilde a -(n-1)\kappa,\tilde b-(n-1)\kappa}\vert \tilde f_0^{\rm III}\rangle\,,\\
\vert  f_n^{\rm IV}\rangle &=& Q^\dagger_{ a,\tilde b} Q^\dagger_{ a-\kappa,\tilde b-\kappa}\dots Q^\dagger_{a -(n-1)\kappa,\tilde b-(n-1)\kappa}\vert  f_0^{\rm IV}\rangle\,,
\eea
In the DBI case we find that four ladder of states have eigenvalues
\bea
c_{\rm I}&=&2(d-1) n(d-2- 2(d-1)n)\,,\\
c_{\rm II} & =& (d-1)(2n-1)(1- 2n(d-1)\,,\\
c_{\rm III}&=& 2(d-1)(n+1)(-2n(d-1)+4-3d)\,,\\
c_{\rm IV} &=& (d-1)(2n+3)(1-2d+2n(d-1))\,,
\eea
which solves the eigenvalue problem with
\bea
{\rm I}&:&\quad\quad\ell=-2(d-1)n,\qquad\qquad\qquad\ell=2(d-1)n+2-d,\label{eq:1l(n)I}\\
{\rm II}&:&\quad\quad\ell=(d-1)(2n-1),\qquad\qquad\ell=-2(d-1)n+1,\label{eq:1l(n)II}\\
\rm III&:&\quad\quad\ell=-2(d-1)n+4-3d,\;\;\quad\ell=2(d-1)(n+1),\label{eq:1l(n)III}\\
{\rm IV}&:&\quad\quad\ell=-(d-1)(3+2n),\qquad\quad\ell=2(d-1)n+2d-1.\label{eq:1l(n)IV}
\eea
which coincides with those found in the main text.

\subsection{P\"osch-Teller potentials}

We will now re-obtain the results of the previous section by making a more direct link with P\"osch-Teller potentials. We first introduce the {\it rapidity} variable $\theta$ defined as
\be
x^{d-1}\equiv \sinh\theta
\ee
and we rescale the perturbation to eliminate the friction term so the multipole equation takes the form
\be
\frac{\dd^2\Phi}{\dd \theta^2}+\left[\frac{d(d-2)}{4(d-1)\sinh^2\theta}+\frac{2}{\cosh^2\theta}\right]\Phi=\left[\frac{2\ell+d-1}{2(d-1)}\right]^2\Phi.
\label{newpot}
\ee
We can now identify this equation with the Schr\"odinger problem with a P\"osch-Teller potential whose Hamiltonian $H_{PT}(\alpha,\beta)$ admits a factorisation in terms of the operators
\bea
A_{\alpha,\beta}&=&\frac{\dd}{\dd\theta}+W_{\alpha,\beta}(\theta),\\
A_{\alpha,\beta}^\dagger&=&-\frac{\dd}{\dd\theta}+W_{\alpha,\beta}(\theta),
\eea
and the super-potentials
\be
\label{eq:PT_potential}
W_{\alpha,\beta}=\alpha\tanh\theta-\frac{\beta}{\tanh\theta},
\ee
for some constants $\alpha$ and $\beta$. We then have
\be
H_{PT}(\alpha,\beta)=A_{\alpha,\beta}^\dagger A_{\alpha,\beta}=-\frac{\dd^2 }{\dd\theta^2}-\frac{\alpha(\alpha+1)}{\cosh^2\theta}+\frac{\beta(\beta-1)}{\sinh^2\theta}+(\alpha-\beta)^2.
\label{HamPT}
\ee
This Hamiltonian admits the following supersymmetric partner:
\be
H^{(s)}_{PT}(\alpha,\beta)=A_{\alpha,\beta}A_{\alpha,\beta}^\dagger=-\frac{\dd^2 }{\dd\theta^2}-\frac{\alpha(\alpha-1)}{\cosh^2\theta}+\frac{\beta(\beta+1)}{\sinh^2\theta}+(\alpha-\beta)^2,
\label{sHamPT}
\ee
which satisfies $H^{(s)}_{PT}(\alpha,\beta)=H_{PT}(-\alpha,-\beta)=H_{PT}(\alpha+1,\beta-1)+4(\alpha-\beta+1)$. If we compare the P\"osch-Teller Hamiltonian \eqref{HamPT} with our equation \eqref{newpot}, we see that we can related both problems upon the identification
\be
\alpha(\alpha+1)=2,\quad\beta(\beta-1)=-\frac{d(d-2)}{4(d-1)}.
\ee
 We can solve these equations for $\alpha$ and $\beta$ to obtain the related P\"osch-Teller potentials: 
 \be
 \alpha_1=1,\quad \alpha_2=-2,\quad \beta_1=\frac{d}{2(d-1)},\quad\beta_2=\frac{d-2}{2(d-1)}.
\ee
We thus have four different P\"osch-Teller Hamiltonians that can be related to our perturbation equations:
\be
H_{\rm I}\equiv H_{PT}(\alpha_1,\beta_1),\quad H_{\rm II}\equiv H_{PT}(\alpha_1,\beta_2),\quad H_{\rm III}\equiv H_{PT}(\alpha_2,\beta_2),\quad
H_{\rm IV}\equiv H_{PT}(\alpha_2,\beta_1)\,.
\ee
In terms of these Hamiltonians, we can write the perturbation equations \eqref{newpot} in the following equivalent forms:
\bea
H_{\rm I}\;\Phi_\ell&=&-\frac{\ell(\ell+d-2)}{(d-1)^2}\Phi_\ell,\\
H_{\rm II}\;\Phi_\ell&=&-\frac{(\ell-1)(\ell+d-1)}{(d-1)^2}\Phi_\ell,\\
H_{\rm III}\;\Phi_\ell&=&-\frac{(\ell+2-2d)(\ell+3d-4)}{(d-1)^2}\Phi_\ell,\\H_{\rm IV}\;\Phi_\ell&=&-\frac{(\ell+1-2d)(\ell+3d-3)}{(d-1)^2}\Phi_\ell.
\eea
The spectrum of the P\"osch-Teller Hamiltonian is given by
\be
E_n(\alpha,\beta)=(\alpha - \beta)^2 - (\alpha - \beta - 2 n)^2=4n(\alpha-\beta-n)
\ee
for integer values of $n$. Using this result for our four related Hamiltonians, we thus find
\bea
&&E_{{\rm I},n}\equiv E_n(\alpha_1,\beta_1)=4n\left(1-\frac{d}{2(d-1)}-n\right)=-\frac{\ell(\ell+d-2)}{(d-1)^2},
\label{eq;EnI}\\
&&E_{{\rm II},n}\equiv E_n(\alpha_1,\beta_2)=2n\left(\frac{d}{d-1}-2n\right)=-\frac{(\ell-1)(\ell+d-1)}{(d-1)^2},\label{eq;EnII}\\
&&E_{{\rm III},n}\equiv E_n(\alpha_2,\beta_2)=2n\frac{6-5d-2(d-1)n}{d-1}=-\frac{(\ell+2-2d)(\ell+3d-4)}{(d-1)^2},\label{eq;EnIII}\\
&&E_{{\rm IV},n}\equiv E_n(\alpha_2,\beta_1)=-4n\left(n+\frac{d}{2(d-1)}+2\right)=-\frac{(\ell+1-2d)(\ell+3d-3)}{(d-1)^2}.\label{eq;EnIV}
\eea
These expressions allow to solve for $\ell$:
\bea
{\rm I}&:&\quad\quad\ell=-2(d-1)n,\qquad\qquad\qquad\ell=2(d-1)n+2-d,\label{eq:1l(n)IPT}\\
{\rm II}&:&\quad\quad\ell=(d-1)(2n-1),\qquad\qquad\ell=-2(d-1)n+1,\label{eq:1l(n)IIPT}\\
\rm III&:&\quad\quad\ell=-2(d-1)n+4-3d,\;\;\quad\ell=2(d-1)(n+1),\label{eq:1l(n)IIIPT}\\
{\rm IV}&:&\quad\quad\ell=-(d-1)(3+2n),\qquad\quad\ell=2(d-1)n+2d-1.\label{eq:1l(n)IVPT}
\eea
From these expressions we see that the second series of I and IV coincide with the multipoles connected by the ladder operators obtained above.

%\vspace{3cm}

%%%%%%%%%%%%%%%%%%%%%%%%%%%%%%%%%
\bibliography{BibKmouflage}
%%%%%%%%%%%%%%%%%%%%%%%%%%%%%%%%%

\end{document}